\documentclass[12pt]{article}
 \usepackage{amssymb}
\usepackage{amsfonts}
 \usepackage{amsmath}
 \usepackage[mathscr]{eucal}
 \usepackage{amsthm}
 \usepackage{bbold}
 \usepackage{bm}
 \usepackage{graphicx}

\def\theequation{\arabic{section}.\arabic{equation}}
\numberwithin{equation}{section}
\newcommand{\be}{\begin{equation}}
\newcommand{\ee}{\end{equation}}
\newcommand{\bea}{\begin{eqnarray}}
\newcommand{\eea}{\end{eqnarray}}

\newcommand{\p}[1]{(\ref{#1})}

\newcounter{rown}

\renewcommand{\theequation}{\thesection.\arabic{equation}}

\begin{document}
\begin{titlepage}
\vspace*{0.1cm}

\begin{center}
{\LARGE\bf Twistorial and space-time descriptions }

\vspace{0.2cm}

{\LARGE\bf  of massless infinite spin}

\vspace{0.2cm}

{\LARGE\bf (super)particles and fields}

\vspace{1.2cm}

{\large\bf I.L. Buchbinder$^{1,2}$,\,\, S. Fedoruk$^3$,\,\,  A.P. Isaev$^{3,4,5}$}

\vskip 1cm

\ $^1${\it Department of Theoretical Physics,
Tomsk State Pedagogical University, \\
634041 Tomsk, Russia}, \\
{\tt joseph@tspu.edu.ru}

\vskip 0.4cm

\ $^2${\it National Research Tomsk State University,\\
634050 Tomsk, Russia}\\

\vskip 0.4cm

\ $^3${\it Bogoliubov Laboratory of Theoretical Physics,
Joint Institute for Nuclear Research, \\
141980 Dubna, Moscow Region, Russia}, \\
{\tt fedoruk@theor.jinr.ru, isaevap@theor.jinr.ru}

\vskip 0.4cm

\ $^4${\it Faculty of Physics, Lomonosov Moscow State University, \\
119991 Moscow, Russia}

\vskip 0.4cm

\ $^5${\it Dubna State University, \\
141982 Dubna, Moscow Region, Russia}

\end{center}

\vspace{0.6cm}

\nopagebreak

\begin{abstract}
\noindent We develop a new twistorial field formulation of a
massless infinite spin particle. Unlike our previous approach
arXiv:1805.09706, the quantization of such a world-line infinite
spin particle model is carried without any gauge fixing. As a
result, we construct a twistorial infinite spin field and derive
its helicity decomposition. Using the field twistor transform, we
construct the space-time infinite (continuous) spin field, which
depends on the coordinate four-vector and additional commuting Weyl
spinor. The equations of motion for infinite spin fields in the
cases of integer and half-integer helicities are derived. We show
that the infinite integer-spin field and infinite half-integer-spin
field form the $\mathcal{N}{=}\,1$ infinite spin supermultiplet. The corresponding
supersymmetry transformations are formulated and their on-shell
algebra is derived. As a result, we find the field realization of the
infinite spin $\mathcal{N}{=}\,1$ supersymmetry.

\end{abstract}

\vspace{0.8cm}

\noindent PACS: 11.10.Ef, 11.30.Cp, 11.30.Pb, 03.65.Pm

\smallskip
\noindent Keywords:   twistors, infinite spin particles, canonical quantization, supersymmetry\\
\phantom{Keywords: }

\newpage

\end{titlepage}
\setcounter{footnote}{0}
\setcounter{equation}{0}

\section{Introduction}

\quad\, Recently, there has been a surge of interest in the description
of particles and fields related to infinite (continuous) spin
representations of the Poincar\'{e} group
\cite{Wigner39,Wigner47,BargWigner}. Although the physical status of
such unitary representations is still not very clear, interest in
them is caused by an identical spectrum of states of the infinite spin
theory \cite{Iv-Mack} and the higher-spin theory
\cite{Vas1989,Vas1991,Vas1992} (see also the reviews
\cite{Vas2001,Vas2002,Vas2005}) and by its potential relation to
the string theory (see \cite{MundSY} and recent paper \cite{Vas2018} and
references in it).
Various problems related to the quantum-mechanical and field descriptions of
such states were considered
in a wide range of works devoted to particles
and fields of an infinite spin (see, e.g.,
\cite{BKRX} -- \cite{Riv18}).

In this paper, we continue to develop an approach to the description of
infinite spin particles and fields, initiated in our previous paper
\cite{BFIR}, where we constructed a new model of an infinite (continuous)
spin particle, which is a generalization of  the twistor formulation of
standard (with fixed helicity) massless particle
\cite{Pen67,PC72,PenRin} to massless infinite spin representations.
Making the quantization of the twistor model with gauge fixing
and using some type of twistor transform, we reproduced these
space-time infinite spin fields, which depend on the position
four-vector and obey the Wigner-Bargmann equations
\cite{Wigner39,Wigner47,BargWigner}.

In  paper  \cite{BFIR}, when quantizing the constructed twistor
model, we imposed partial gauge fixings for gauge symmetries
generated by some first class constraints. This yielded after
quantization ``limited'' fields describing the massless
representations of the infinite spin. Now we will carry out the
quantization of the twistor model without any gauge fixing and use a
different field twistor transform. As a result of such a
quantization procedure, we obtain infinite spin fields that have a
more transparent decomposition into helicities in the twistor
formulation. Besides, we show that our twistorial model reproduces
the space-time--spinorial formulation of infinite spin fields with
integer helicities proposed in \cite{BuchKrTak}. In addition, using
the field twistor transform, we can now get the
space-time--spinorial formulation of infinite spin fields with
half-integer helicities. Moreover, such a formulation allows us to
construct a supermultiplet of infinite spins \cite{BKRX} (see also
the recent development in \cite{Zin}) and describe some of its
properties. It is worth pointing out that the aspects of infinite
spin supersymmetry have almost been unconsidered in the literature
earlier.

The plan of this paper is as follows. In Sect.\,2, we describe the
world-line twistor formulation of the infinite (continuous) spin
particle, which we built in  \cite{BFIR}. We present the twistorial
constraints of the model and coordinate twistor transform to
space-time formulation. In Sect.\,3, by using canonical
transformation, we introduce suitable phase variables into the
twistor formulation, in which all twistor constraints take a very
simple form appropriate for quantization. We use the Dirac
scheme of canonical quantization of gauge systems, where the first
class constraints are imposed on the wave functions. As a
result, we obtain the field equations of motion and the twistor field of infinite spin
is a solution of these equations. The resulting
twistor field has the $\mathrm{U}(1)$ charge that corresponds to the
method of field description of the infinite spin particle. Integer
or half-integer values of this charge correspond to infinite spin
particles with integer or half-integer values of helicities. Also,
the decomposition of the twistor field in an infinite number of
states with fixed helicities is found. In Sect.\,4, we construct a
field twistor transform, which determines the space-time field of
the infinite (continuous) spin particle, according to the twistor
field obtained in the previous section. These space-time fields
depend, in addition to the four-vector coordinates, also on the
components of the additional commuting Weyl spinor. We have found
the equations of motion for the fields of infinite spin particles,
both in the case of integer and half-integer helicities. In
Sect.\,5, the $\mathcal{N}{=}\,1$ infinite spin supermultiplet is
constructed. On the mass-shell this supermultiplet consists of two
complex infinite spin fields with integer and half-integer
helicities, respectively. In Sect.\,6, we give some comments on the
results obtained. In two Appendices, we give the proof of the
equations of motion for the space-time infinite spin fields and
present the explicit form of the momentum wave function in the
space-time formulation.

\setcounter{equation}{0}

\section{Twistorial formulation of infinite spin particle}

\quad\, In \cite{BFIR}, we constructed the twistorial formulation of the infinite (continuous) spin particle.
It is described\footnote{As in \cite{BFIR} we use the following notation. The space-time
metric is $\eta_{mn}={\rm diag}(+1,-1,-1,-1)$. The totally antisymmetric
tensor $\epsilon_{mnkl}$ has the component $\epsilon_{0123}=-1$.
The two-component Weyl spinor indices are raised and lowered by
$\epsilon_{\alpha\beta}$, $\epsilon^{\alpha\beta}$,
$\epsilon_{\dot\alpha\dot\beta}$, $\epsilon^{\dot\alpha\dot\beta}$
with the nonvanishing components
$\epsilon_{12}=-\epsilon_{21}=\epsilon^{21}=-\epsilon^{12}=1$:
$\psi_\alpha=\epsilon_{\alpha\beta}\psi^\beta$,
$\psi^\alpha=\epsilon^{\alpha\beta}\psi_\beta$, etc. Relativistic
$\sigma$-matrices are $ (\sigma_m)_{\alpha\dot\beta}=({\bf
1_2};\sigma_1,\sigma_2,\sigma_3)_{\alpha\dot\beta} $, where
$\sigma_1,\sigma_2,\sigma_3$ are the Pauli matrices. The matrices $
(\tilde\sigma_{m})^{\dot\alpha\beta}=\epsilon^{\dot\alpha\dot\delta}\epsilon^{\beta\gamma}
(\sigma_m)_{\gamma\dot\delta}=({\bf
1_2};-\sigma_1,-\sigma_2,-\sigma_3)^{\dot\alpha\beta} $ satisfy $
\sigma^m_{\alpha\dot\gamma}\tilde\sigma^{n\,\dot\gamma\beta}+\sigma^m_{\alpha\dot\gamma}\tilde\sigma^{n\,\dot\gamma\beta}
=2\,\eta^{mn}\delta^\beta_\alpha$ and
$\sigma^m_{\alpha\dot\beta}\tilde\sigma_n^{\dot\beta\alpha}=2\,\delta^m_n$.
The link between the Minkowski four-vectors and spinorial quantities is
given by
$A_{\alpha\dot\beta}={\textstyle\frac{1}{\sqrt{2}}}\,A_m(\sigma^m)_{\alpha\dot\beta}$,
$A^{\dot\alpha\beta}={\textstyle\frac{1}{\sqrt{2}}}\,A_m(\tilde\sigma^m)^{\dot\alpha\beta}$,
$A_m
={\textstyle\frac{1}{\sqrt{2}}}\,A_{\alpha\dot\beta}(\tilde\sigma_m)^{\dot\beta\alpha}$,
so that $ A^m B_m =A_{\alpha\dot\beta}B^{\dot\beta\alpha}$.}
by the bosonic Weyl spinors  $\pi$, $\bar{\pi}$,
$\rho$, $\bar{\rho}$ and their canonically conjugated spinors
$\omega$, $\bar{\omega}$, $\eta$, $\bar{\eta}$ with the components
 \begin{equation}
 \label{nov01}
\pi_{\alpha} \; , \;\;\; \bar\pi_{\dot\alpha}:=(\pi_{\alpha})^* \; , \;\;\;
\rho_{\alpha} \; , \;\;\; \bar\rho_{\dot\alpha}:=(\rho_{\alpha})^*\; , \;\;\; \omega^{\alpha}\; , \;\;\; \bar \omega^{\dot\alpha}:=(\omega^{\alpha})^* \; , \;\;\;\eta^{\alpha} \; , \;\;\; \bar{\eta}^{\dot\alpha}:=(\eta^{\alpha})^* \; .
 \end{equation}
The nonzero Poisson brackets of these spinors are
\begin{equation}
\label{tw-PB-sp}
\left\{ \omega^{\alpha}, \pi_{\beta} \right\}=\left\{ \eta^{\alpha}, \rho_{\beta} \right\}=\delta^\alpha_\beta\,,\qquad
\left\{\bar \omega^{\dot\alpha}, \bar\pi_{\dot\beta} \right\}=\left\{\bar \eta^{\dot\alpha}, \bar\rho_{\dot\beta} \right\}=\delta^{\dot\alpha}_{\dot\beta}\,.
\end{equation}
 We assume that all spinors are functions of a time parameter $\tau$.
Twistorial Lagrangian of the infinite (continuous) spin particle is
written in the form \cite{BFIR}:
\begin{equation}
\label{L-tw}
{\cal L}_{twistor}= \pi_{\alpha}\dot \omega^{\alpha} \
+ \ \bar\pi_{\dot\alpha}\dot{\bar \omega}^{\dot\alpha}  \ + \
\rho_{\alpha}\dot \eta^{\alpha}  \ + \
\bar\rho_{\dot\alpha}\dot{\bar \eta}^{\dot\alpha} \ + \
l\,{\mathcal{M}} \ + \  k\,{\mathcal{U}} \ + \  \ell\,{\mathcal{F}}
\ +\  \bar\ell\,\bar{\mathcal{F}}\,,
\end{equation}
where $l(\tau)$, $k(\tau)$, $\ell(\tau)$, $\bar\ell(\tau)$
are the Lagrange multipliers for the constraints
\begin{eqnarray}
\label{M-constr-def}
\mathcal{M} &:=& \pi^{\alpha}\rho_{\alpha}\,\bar\rho_{\dot\alpha}\bar\pi^{\dot\alpha}-M^2\ \approx\  0\,,
\\ [6pt]
\label{constr-tw}
{\mathcal{F}}&:=& \eta^{\alpha}\pi_{\alpha}-1\approx 0\,,\qquad\quad
\bar{\mathcal{F}}\ :=\ \bar\pi_{\dot\alpha}\bar \eta^{\dot\alpha}-1\approx 0\,,
\\ [6pt]
\label{const-tw-4}
{\mathcal{U}}& :=& i \; (\omega^{\alpha}\pi_{\alpha}-\bar\pi_{\dot\alpha}\bar \omega^{\dot\alpha}
+\eta^{\alpha}\rho_{\alpha}-\bar\rho_{\dot\alpha}\bar \eta^{\dot\alpha})
\approx 0\,,
\end{eqnarray}
and we use standard notation $\dot{x}(\tau) := \partial_\tau x(\tau)$
for any function of $\tau$.
The real constant $M$ in \p{M-constr-def} is
represented in the form
\begin{equation}
\label{M-mu}
M^2=\mu^2/2\,,
\end{equation}
where the dimensionful parameter $\mu$ fixes the irreducible representation
of the Poincare group describing massless infinite (continuous) spin particles (see
(\ref{P2W2-sp}) below). This constant $\mu$ has nothing to do with the mass parameter
in the twistor formulations of massive particles with spin (see,
for example, \cite{FZ03,FFLM06,FL14,AFIL,IsP} and references therein).
One can check that four first class
constraints \eqref{M-constr-def}, \eqref{constr-tw}, \eqref{const-tw-4}
generate abelian Lie algebra with respect to the twistorial Poisson brackets  \eqref{tw-PB-sp}. The Lie group ${\cal H}$ corresponding to
this abelian Lie algebra acts in the space of spinors (\ref{nov01}) as follows:
 \begin{eqnarray}
 \label{gauget0}
  \left(\!\!
 \begin{array}{cc}
 \pi_1 \! & \! \rho_1 \\
 \pi_2 \! & \! \rho_2
 \end{array}
 \!\!\right)
 & \rightarrow &
 \left(\!\!
 \begin{array}{cc}
 \pi_1 \! & \! \rho_1 \\
 \pi_2 \! & \! \rho_2
 \end{array}
\! \!\right)  \left(\!\!
 \begin{array}{cc}
 e^{i\beta} \! & \! \alpha \\
 0 \! & \! e^{i\beta}
 \end{array}
 \!\!\right) \, , \;\;  \\ [6pt]
 \label{gauget}
 \left(\!
 \begin{array}{cc}
 \eta_1 \! & \! \omega_1 \\
 \eta_2 \! & \! \omega_2
 \end{array}
 \!\!\right)
 & \rightarrow &  \left(\!\!
 \begin{array}{cc}
 \eta_1 \! & \! \omega_1 \\
 \eta_2 \! & \! \omega_2
 \end{array}
 \!\!\right)  \left(\!\!
 \begin{array}{cc}
 e^{-i\beta} \! & \! -\alpha \\
 0 \! & \! e^{-i\beta}
 \end{array}
 \!\!\right) +  M^{-2}(\bar\rho_{\dot\alpha}\bar\pi^{\dot\alpha})  \left(\!\!
 \begin{array}{cc}
 \pi_1 \! & \! \rho_1 \\
 \pi_2 \! & \! \rho_2
 \end{array}
 \!\!\right) \left(\!\!
 \begin{array}{cc}
 \gamma  \! & \! 0 \\
 0 \! & \! -\gamma
 \end{array}
 \!\!\right)  , \\ [6pt]
 && {plus} \;\; {complex} \;\; {conjugated} \;\; {transformations} \; ,
 \nonumber
\end{eqnarray}
where $\beta(\tau) , \gamma(\tau) \in \mathbb{R}$ and
$\alpha(\tau) \in \mathbb{C}\backslash 0$ are the parameters of
the gauge transformations
\begin{equation}
\label{PB-g-tr}
X\quad \rightarrow \quad e^{\,\displaystyle A}* X:=
X+\sum\limits_{n=1}^{\infty}\frac{1}{n!}\,\underbrace{ \{A, \{A,\cdots \{ A \,,X \}\cdots \} \}}_{n\, {\rm times}}\,,
\end{equation}
for arbitrary phase variable $X$, where
\begin{equation}
\label{A-g-tr}
A \ := \ \alpha\,{\mathcal{F}}+\bar\alpha\,\bar{\mathcal{F}}+ \beta\,{\mathcal{U}}-M^{-2}\gamma\,{\mathcal{M}}\,.
\end{equation}
Constraints \eqref{M-constr-def}, \eqref{constr-tw}, \eqref{const-tw-4} eliminate
(if we also add four gauge fixing conditions) eight degrees of freedom
 from sixteen degrees in (\ref{nov01}).

The space-time (Wigner-Bargmann) formulation of the infinite (continuous) spin particle is
given in the phase space
with two four-vectors $x_m\leftrightarrow x_{\alpha\dot\alpha}$,
 $y_m\leftrightarrow y_{\alpha\dot\alpha}$ and their conjugated momenta
$p_m \leftrightarrow p_{\alpha\dot\alpha}$,
$q_m \leftrightarrow q_{\alpha\dot\alpha}$
with the Poisson brackets
\begin{equation}
\label{PB-sp}
\left\{ x^{\dot\alpha\alpha}, p_{\beta\dot\beta} \right\}=\delta^{\alpha}_{\beta}\delta^{\dot\alpha}_{\dot\beta}\,,\qquad
\left\{ y^{\dot\alpha\alpha}, q_{\beta\dot\beta} \right\}=\delta^{\alpha}_{\beta}\delta^{\dot\alpha}_{\dot\beta}\,.
\end{equation}
The link of twistorial and space-time formulations  is carried out by the generalization of the Cartan-Penrose relations \cite{Pen67,PC72,PenRin}
\begin{equation}
\label{p-resol}
p_{\alpha\dot\beta}=\pi_{\alpha}\bar\pi_{\dot\beta}\,,
\end{equation}
\begin{equation}
\label{w-resol}
q_{\alpha\dot\beta}=\pi_{\alpha}\bar\rho_{\dot\beta}+\rho_{\alpha}\bar\pi_{\dot\beta}\,,
\end{equation}
and by the following incidence relations:
\begin{equation}
\label{y-tw}
\omega^{\alpha}=\bar\pi_{\dot\alpha}x^{\dot\alpha\alpha}+\bar\rho_{\dot\alpha}y^{\dot\alpha\alpha}\,,\qquad
\bar \omega^{\dot\alpha}=x^{\dot\alpha\alpha}\pi_{\alpha}+y^{\dot\alpha\alpha}\rho_{\alpha}\,,
\end{equation}
\begin{equation}
\label{z-tw}
\eta^{\alpha}=\bar\pi_{\dot\alpha}y^{\dot\alpha\alpha}\,,\qquad
\bar \eta^{\dot\alpha}=y^{\dot\alpha\alpha}\pi_{\alpha}\,.
\end{equation}
The space-time formulation of the Lagrangian for the infinite (continuous) spin particle
in the first-order formalism is
\begin{equation}
\label{L-sp}
{\cal L}_{sp.-time}(\tau) = p_m \, \dot x^m + q_m \, \dot y^m + e \, T +
e_1 \, T_1 + e_2 \, T_2 + e_3 \, T_3\,.
\end{equation}
Here the functions $e(\tau)$, $e_1(\tau)$, $e_2(\tau)$, $e_3(\tau)$
are the Lagrange multipliers for constraints
\cite{Wigner47,BargWigner}:
\begin{eqnarray}
T &:=& p_m p^m \ \approx \ 0 \,, \label{const-sp}\\
T_1 &:=& p_m q^m \ \approx \ 0  \,,  \label{const-sp-1}\\
T_2 &:=& q_m q^m +\mu^2 \ \approx \ 0  \,,  \label{const-sp-2}\\
T_3 &:=& p_m y^m -1 \ \approx \ 0 \,.  \label{const-sp-3}
\end{eqnarray}
Note that pairs of
spinors $\pi_{\alpha}$, $\bar\omega^{\dot\alpha}$ and $\rho_{\alpha}$, $\bar \eta^{\dot\alpha}$
form two Penrose twistors
\begin{equation}
\label{Z-tw-def}
Z_{A} := \left( \pi_\alpha, \bar{\omega}^{\dot\alpha} \right) , \qquad
Y_{A} := \left( \rho_\alpha, \bar{\eta}^{\dot\alpha} \right) .
\end{equation}
Conjugated spinors $\bar\pi_{\dot\alpha}$, $\omega^{\alpha}$ and
 $\bar\rho_{\dot\alpha}$, $\eta^{\alpha}$
constitute the dual twistors
\begin{equation}
\label{Y-tw-def}
\bar Z^{A}:=
\left(\!
\begin{array}{c}
\omega^{\alpha} \\
-\bar{\pi}_{\dot{\alpha}} \\
\end{array}
\!\right) \; ,
\qquad
\bar Y^{A}:=
\left(\!
\begin{array}{c}
\eta^{\alpha} \\
-\bar{\rho}_{\dot{\alpha}} \\
\end{array}
\!\right) \; .
\end{equation}
So the description of infinite spin particles uses with necessity a couple of twistors
as opposed to the one-twistor description of the massless particle with fixed helicity.

On the shell of constraints  \eqref{M-constr-def}-\eqref{const-tw-4},
 (\ref{M-mu}) (or on the shell of constraints  \eqref{const-sp}-\eqref{const-sp-3})
we have got the conditions for the Casimir operators
\begin{equation}
\label{P2W2-sp}
p_m p^m\approx 0\,, \qquad W_m W^m \approx -\mu^2
\end{equation}
where
\begin{equation}
\label{W-def}
W_m=\frac12\,\varepsilon_{mnkl}p^n M^{kl}  \, ,
\end{equation}
are the components of the Pauli-Luba\'{n}ski pseudovector and $p_m$ and $M_{kl} :=
(x_k p_l - x_l p_k) + (y_k q_l - y_l q_k)$ are the Poincar\'{e} algebra generators.
The models with Lagrangians \p{L-tw} and \eqref{L-sp} are equivalent
at the classical level (see \cite{BFIR})
and describe massless particles with the infinite (continuous) spin.

Following \cite{Pen67},\cite{PC72},\cite{PenRin}
we choose the norms of twistors \eqref{Z-tw-def}, \eqref{Y-tw-def}
as
\begin{equation}
\label{norm-tw}
\bar Z^{A}Z_A\, =\, \omega^{\alpha}\pi_{\alpha}-\bar\pi_{\dot\alpha}\bar \omega^{\dot\alpha}\,,\qquad
\bar Y^{A}Y_A\, =\, \eta^{\alpha}\rho_{\alpha}-\bar\rho_{\dot\alpha}\bar \eta^{\dot\alpha}\,,
\end{equation}
and write the constraint \eqref{const-tw-4} in concise form
\begin{equation}
\label{const-tw-4-tw}
{\mathcal{U}}\, = \, i \, (\bar Z^{A}Z_A+\bar Y^{A}Y_A) \,\approx\, 0\,.
\end{equation}
The norm $\bar Z^{A}Z_A$
of the twistor $Z$ commutes with constraints \eqref{M-constr-def},
\eqref{constr-tw}, \eqref{const-tw-4} and therefore
is independent of $\tau$. For a massless particle with fixed helicity
the norm $\bar Z^{A}Z_A$ defines the helicity operator
\begin{equation}
\label{hel-tw}
\Lambda=\frac{i}{2}\,\bar Z^{A}Z_A\,.
\end{equation}
So in the considered model of the infinite (continuous) spin particle,
in view of the constraint \eqref{const-tw-4-tw},
the particle helicity is not fixed since it is proportional to
$-\bar Y^{A}Y_A$.

Let us make some comments on the role of the constraints \p{constr-tw}.
First of all, we recall the statement from  \cite{BFIR} that the matrices
\begin{equation}
\label{Lor-matr}
\parallel\! A_\alpha{}^{b}\!\parallel \ := \ \frac{1}{\sqrt{\pi\rho}} \left(\!\!
 \begin{array}{cc}
 \pi_1 \! & \! \rho_1 \\
 \pi_2 \! & \! \rho_2
 \end{array}
 \!\!\right),
\end{equation}
where $\pi\rho:=\pi^\alpha\rho_\alpha$ can be considered as elements of
$SL(2,\mathbb{C})$.
Moreover, the Lorentz transformation
\begin{equation}
\label{k-p}
A \, \hat k \, A^+ = \hat p\,,
\end{equation}
with $A \in SL(2,\mathbb{C})$ given in \p{Lor-matr} converts
test massless momentum
\begin{equation}
\label{k-matr}
\hat k \ := \ \parallel\! k_{a\dot b}\!\parallel \ := \ \left(\!\!
 \begin{array}{cc}
 \sqrt{2}k \! & \! 0 \\
 0 \! & \! 0
 \end{array}
 \!\!\right),\qquad
k^m=(k;0,0,k)
\end{equation}
to momentum   \p{p-resol}
\begin{equation}
\label{p-matr}
\hat p \ := \ \parallel\! p_{\alpha\dot \beta}\!\parallel \ := \ \left(\!\!
 \begin{array}{cc}
 \pi_{1}\bar\pi_{\dot 1}  &  \pi_{1}\bar\pi_{\dot 2} \\
 \pi_{2}\bar\pi_{\dot 1}  &  \pi_{2}\bar\pi_{\dot 2}
 \end{array}
 \!\!\right),
\end{equation}
where we assume $|\pi\rho|=\sqrt{2}k$.
Inverse Lorentz transformations that translate $\hat p$
into the basis of test momentum $\hat k$
are given by the matrix
\begin{equation}
\label{Lor-matr-1}
A^{-1} \ := \ \parallel\! (A^{-1}){}_a{}^{\beta}\!\parallel \ := \ \frac{1}{\sqrt{\pi\rho}} \left(\!\!
 \begin{array}{cc}
 -\rho^1 \! & \! -\rho^2  \\
 \pi^1 \! & \! \pi^2
 \end{array}
 \!\!\right) .
\end{equation}
In the twistor realization  the Pauli-Luba\'{n}ski pseudovector
with components \p{W-def} has the form
\begin{equation}
\label{PL-tw}
W_{\alpha\dot\beta}=
\Lambda \cdot p_{\alpha\dot\beta}
-\frac{i}{2}\,\Big[(\bar\pi_{\dot\gamma}\bar\eta^{\dot\gamma})\pi_{\alpha}\bar\rho_{\dot\beta}
-(\pi_{\gamma}\eta^{\gamma})\rho_{\alpha}\bar\pi_{\dot\beta} \Big]
+\frac{i}{2}\,\Big[(\bar\pi^{\dot\gamma}\bar\rho_{\dot\gamma})\pi_{\alpha}\bar\eta_{\dot\beta}- (\pi^{\gamma}\rho_{\gamma})\eta_{\alpha}\bar\pi_{\dot\beta}\Big]\,,
\end{equation}
where $p_{\alpha\dot\beta}$ is defined in \p{p-resol}.
Therefore, this pseudovector in the basis of test momentum is defined by the relation
\begin{equation}
\label{PL-tw-0}
\stackrel{\mathrm{o}}{W}_{a\dot b}=
(A^{-1}){}_a{}^{\alpha}\, W_{\alpha\dot\beta}\, (A^{-1 +}){}^{\dot\beta}{}_{\dot b}
\end{equation}
and has the components
\begin{equation}
\label{PL-tw-0c}
\stackrel{\mathrm{o}}{W}_{1\dot 1}=|\pi\rho|\,V\,,\qquad
\stackrel{\mathrm{o}}{W}_{2\dot 2}=0\,,\qquad
\stackrel{\mathrm{o}}{W}_{1\dot 2}=-2i|\pi\rho|\,(\bar\pi_{\dot\alpha}\bar\eta^{\dot\alpha})\,,\qquad
\stackrel{\mathrm{o}}{W}_{2\dot 1}=2i|\pi\rho| \,(\eta^{\alpha}\pi_{\alpha})\,,
\end{equation}
where (cf. (\ref{const-tw-4}))
\begin{equation}
\label{PL-tw-0c-def}
V:= \frac{i}{2}\left(\omega^{\alpha}\pi_{\alpha} -\bar\pi_{\dot\alpha}\bar\omega^{\dot\alpha}-
\eta^{\alpha}\rho_{\alpha} +\bar\rho_{\dot\alpha}\bar\eta^{\dot\alpha}\right)=
\Lambda-\frac{i}{2}\left(\eta^{\alpha}\rho_{\alpha} -\bar\rho_{\dot\alpha}\bar\eta^{\dot\alpha}\right)\,,
\end{equation}
Using the standard analysis of the massless representations of the Poincar\'{e} group,
we find from expressions  \p{PL-tw-0c} that the quantities
\begin{equation}
\label{E2-gen}
V\,,\qquad
E=E_1+iE_2 := \sqrt{2}{M}\,(\bar\pi_{\dot\alpha}\bar\eta^{\dot\alpha})\,,\qquad
\bar E=E_1-iE_2 := \sqrt{2}{M}\,(\eta^{\alpha}\pi_{\alpha})
\end{equation}
are the generators of the small subgroup $E(2)$.
The generators $E$ and $\bar E$ yield translations
in $\mathbb{C}$ which correspond
to the parameter $\alpha$ in \p{gauget0}, \p{gauget}. Then, constraints \p{constr-tw}
fix the Casimir operator of the small subgroup,
\begin{equation}
\label{E2-Cas}
E\bar E=(E_1)^2+(E_2)^2= 2{M}^2=\mu^2\,,
\end{equation}
and therefore fix the Casimir operator $W^2$ of the Poincar\'{e} group.

\setcounter{equation}{0}
\section{Alternative quantization of twistorial model and infinite spin twistor field}

\quad\, The twistorial model with Lagrangian (\ref{L-tw}) possesses
the gauge symmetry under transformations (\ref{gauget}) generated by
the first class constraints \eqref{M-constr-def},
\eqref{constr-tw} and \eqref{const-tw-4}.
In \cite{BFIR}, we performed the quantization of the model  \p{L-tw} after
partial fixing the gauge as $s=1=\bar{s}$ (the definition of
$s,\bar{s}$ is given below).
In this paper, we quantize the model on the base of a more general procedure without any gauge fixing.

To simplify the quantization
 of the model, in \cite{BFIR} we made a canonical transformation
 of spinorial variables (\ref{nov01}) which is
 motivated by the gauge transformation (\ref{gauget0}), (\ref{gauget}).
 For variables $\pi_{\alpha}, \; \rho_{\alpha}$ we have
\begin{equation}
 \label{gaugei}
  \begin{array}{c}
 \left(\!\!
 \begin{array}{cc}
 \pi_1 \! & \! \rho_1 \\
 \pi_2 \! & \! \rho_2
 \end{array}
 \!\!\right) 
 = \sqrt{M} \left(\!\!
 \begin{array}{cc}
 p^{(z)}_1 \! & \! 0 \\
p^{(z)}_2 \! &  p^{(s)}/p^{(z)}_1
 \end{array}
\! \!\right)  \left(\!\!
 \begin{array}{cc}
 1 \! & \! p^{(t)} \\
 0 \! & \! 1
 \end{array}
 \!\!\right) \, , \;\;
   \end{array}
 \end{equation}
 where new variables are
\begin{equation}
\label{def-new-var-z}
p^{(z)}_{\alpha} = \pi_{\alpha}/\sqrt{M}\,,\qquad
p^{(s)}=\pi^{\alpha}\rho_{\alpha}/M\,,\qquad
p^{(t)}=\rho_{1}/\pi_{1}
\end{equation}
For the conjugated momentum variables we have
\begin{equation}
\label{def-new-var-bz}
\bar\pi_{\dot\alpha}, \; \bar\rho_{\dot\alpha} \;\;\;\; \to \;\;\;\;
\bar p^{(z)}_{\dot\alpha}= \bar\pi_{\dot\alpha}/\sqrt{M}\,,\qquad
\bar{p}^{(s)}=\bar\rho_{\dot\alpha}\bar\pi^{\dot\alpha}/M\,,\qquad
\bar{p}^{(t)}=\bar\rho_{1}/\bar\pi_{1}\, .
\end{equation}
The corresponding new coordinate variables
$z^{\alpha}$, $s$, $t$
are connected with old twistorial variables by means of
relations \cite{BFIR} which also motivated by gauge transformations
(\ref{gauget})
\begin{equation}
\label{gaugej}
 \left(\!\!
 \begin{array}{cc}
 \eta_1 \! & \! \omega_1 \\
 \eta_2 \! & \! \omega_2
 \end{array}
 \!\!\right) 
 = \left(\!\!
 \begin{array}{cc}
 0 \! & \!  z_1/\sqrt{M} \\
 -t/\pi_1 \! &   z_2/\sqrt{M}
 \end{array}
\! \!\right)  \left(\!\!
 \begin{array}{cc}
 1 \! & \! - p^{(t)} \\
 0 \! & \! 1
 \end{array}
 \!\!\right) +
 \frac{s}{M}  \left(\!\!
 \begin{array}{cc}
 \pi_1 \! & \! \rho_1 \\
 \pi_2 \! & \! \rho_2
 \end{array}
 \!\!\right)   \left(\!\!
 \begin{array}{cc}
 1 \! & \! 0 \\
 0 \! & \! -1
 \end{array}
 \!\!\right)\, , \;\;
 \end{equation}
where
\begin{eqnarray}\label{mu-can}
\omega^\alpha &=&  \,\frac{1}{\sqrt{M}}\,z^{\alpha} \ - \
\frac{1}{M}\,s\,\rho^{\alpha}  \ - \
\frac{\delta^{\alpha 1}}{\pi_1}\, t\, p^{(t)}  \,,\\
\eta^\alpha &=&  \ \,
\frac{1}{M}\,s\,\pi^{\alpha}  \ + \ \frac{\delta^{\alpha 1}}{\pi_1}\, t  \, . \label{omega-can}
\end{eqnarray}
By complex conjugation we obtain from (\ref{mu-can}), (\ref{omega-can})
 the relations for conjugated coordinates
 $\bar{z}^{\dot\alpha}$, $\bar{s}$, $\bar{t}$.
The nonzero Poisson brackets, which are consistent with
 (\ref{tw-PB-sp}), are
\begin{equation}
\label{PB-n-var}
\begin{array}{c}
\left\{ z^{\alpha}, p^{(z)}_{\beta} \right\}= \delta^\alpha_\beta\,, \qquad
\left\{\bar z^{\dot\alpha}, \bar p^{(z)}_{\dot\beta} \right\}=
\delta^{\dot\alpha}_{\dot\beta}\,,\\ [7pt]
\left\{ s, p^{(s)} \right\}= \left\{\bar s, \bar p^{(s)} \right\}=1\,,\qquad
\left\{ t, p^{(t)} \right\}= \left\{\bar t, \bar p^{(t)} \right\}=1\, .
\end{array}
\end{equation}

In terms of new variables the constraints \eqref{M-constr-def}, \eqref{constr-tw}, \eqref{const-tw-4} take, up to multipliers, the form
\begin{eqnarray}
\label{M-constr-def-n}
{\mathcal{M}}^\prime&:=& p^{(s)} \bar{p}^{(s)} -1\ \approx\  0\,,
\\ [6pt]
\label{constr-tw-n}
{\mathcal{F}}^\prime&:=& t - 1\ \approx\ 0\,,\qquad \bar{\mathcal{F}}^\prime \ := \ \bar{t} - 1\ \approx\  0\,,
\\ [6pt]
\label{const-tw-4-n}
{\mathcal{U}}^\prime&:=& \frac{i}{2}\,\left(z^{\alpha}p^{(z)}_{\alpha} - \bar z^{\dot\alpha}\bar p^{(z)}_{\dot\alpha}\right) \ +
\ i \left( s p^{(s)} - \bar s\bar p^{(s)} \right)
\ \approx\ 0\,,
\end{eqnarray}
and we write twistorial Lagrangian \p{L-tw} as
\begin{equation}
\label{L-tw-n}
{\cal L}^\prime \ = \
p^{(z)}_{\alpha}\dot z^{\alpha} \ + \ \bar p^{(z)}_{\dot\alpha}\dot{\bar z}^{\dot\alpha}  \ + \
p^{(s)}\dot s  \ + \  \bar p^{(s)}\dot{\bar s} \ + \
p^{(t)}\dot t  \ + \  \bar p^{(t)}\dot{\bar t} \ + \
l^\prime\,{\mathcal{M}}^\prime \ + \  k^\prime\,{\mathcal{U}}^\prime \ + \
\ell^\prime\,{\mathcal{F}}^\prime \ +\  \bar\ell^\prime\,\bar{\mathcal{F}}^\prime\,,
\end{equation}
where $l^\prime(\tau)$, $k^\prime(\tau)$, $\ell^\prime(\tau)$, $\bar\ell^\prime(\tau)$ are the Lagrange multipliers
for the constraints \eqref{M-constr-def-n}, \eqref{constr-tw-n}, \eqref{const-tw-4-n}.
Gauge transformations \eqref{gauget0}, \eqref{gauget} of new phase variables \eqref{PB-n-var}
have the form
\begin{equation}
\label{trans-tw-n}
\begin{array}{l}
z^{\alpha}\,\rightarrow\,e^{-i\beta}z^{\alpha}  \,,\qquad
s\,\rightarrow\,e^{-2i\beta}s +\gamma \bar p^{(s)}\,,\qquad
t\,\rightarrow\,t  \,, \\ [6pt]
p^{(z)}_{\alpha}\,\rightarrow\,e^{i\beta}p^{(z)}_{\alpha}\,,\qquad
p^{(s)}\,\rightarrow\,e^{2i\beta}p^{(s)}  \,,\qquad
p^{(t)}\,\rightarrow\,p^{(t)}+\alpha
\end{array}
\end{equation}
plus complex conjugated transformations.
Since the constraints \eqref{M-constr-def-n}, \eqref{constr-tw-n}, \eqref{const-tw-4-n} form
abelian algebra with respect to the Poisson brackets, gauge transformations of the Lagrange multipliers
have the simple form
\begin{equation}
\label{trans-lm-n}
l^\prime\,\rightarrow\,l^\prime-\dot\gamma\,,\qquad   k^\prime\,\rightarrow\,k^\prime+2\dot\beta\,,\qquad
\ell^\prime\,\rightarrow\,\ell^\prime+\dot\alpha \,,\qquad  \bar\ell^\prime\,\rightarrow\,\bar\ell^\prime+\dot{\bar\alpha}\,.
\end{equation}

The important fact is that in
terms of the new variables all constraints \eqref{M-constr-def-n}, \eqref{constr-tw-n}, \eqref{const-tw-4-n}
acquire a simpler form which is suitable
 for quantization $[.,.] = i \, \{.,.\} $.
We will perform quantization in the momentum representation,
where the operators of the dynamical variables are realized as follows (we take $\hbar=1$):
\begin{equation}
\label{realiz-var}
\begin{array}{l}
{\displaystyle \hat z^{\alpha}=i\frac{\partial}{\partial p^{(z)}_{\alpha}}\,, \quad
\hat p^{(z)}_{\alpha} =p^{(z)}_{\alpha} \,,\qquad
\hat {\bar z}^{\dot \alpha}=
i\frac{\partial}{\partial {\bar p}^{(z)}_{\, \dot \alpha} }
\,, \quad \hat {\bar p}^{(z)}_{\, \dot \alpha} = {\bar p}^{(z)}_{\, \dot \alpha}
\,, } \\ [9pt]
{\displaystyle \hat s= i\frac{\partial}{\partial p^{(s)}} \,,
\quad \hat p^{(s)} =  p^{(s)}\,,\qquad
\hat {\bar s} = i\frac{\partial}{\partial \bar p^{(s)}}\,, \quad
 \hat {\bar p}^{(s)} = {\bar p}^{(s)}\,, }\\ [9pt]
{\displaystyle \hat t= i\frac{\partial}{\partial p^{(t)}}\,,
\quad \hat p^{(t)} = p^{(t)} \,,\qquad
\hat {\bar t} = i\frac{\partial}{\partial {\bar p}^{(t)} } \,,
\quad \hat {\bar p}^{(t)} = {\bar p}^{(t)} \,. }
\end{array}
\end{equation}
The corresponding wave function has the form
\begin{equation}
\label{wf-1}
\Psi=\Psi\big( p^{(z)}_{\alpha},\, \bar{p}^{(z)}_{\dot\alpha}; \,
p^{(s)}, \, \bar{p}^{(s)}; \, p^{(t)}, \, \bar{p}^{(t)}\big)\,.
\end{equation}
The wave function \p{wf-1}  describes physical states and obeys the equations
\begin{eqnarray}
\label{M-constr-def-q}
&& \left(p^{(s)} \bar{p}^{(s)} -1\right) \Psi^{(c)} \ =\  0\,,
\\ [6pt]
\label{constr-tw-q}
&& \frac{\partial}{\partial p^{(t)}}\ \Psi^{(c)}\ =\
\frac{\partial}{\partial \bar{p}^{(t)}} \ \Psi^{(c)}\ =\ -i\,\Psi^{(c)}\,,
\\ [6pt]
\label{const-tw-4-qq}
&& \left[\frac{1}{2}\,\left(p^{(z)}_{\alpha}\frac{\partial}{\partial p^{(z)}_{\alpha}} \ -\
\bar{p}^{(z)}_{\dot\alpha}\frac{\partial}{\partial \bar{p}^{(z)}_{\dot \alpha}}\right) \ +\
p^{(s)} \frac{\partial}{\partial p^{(s)}} \ - \
\bar{p}^{(s)}\frac{\partial}{\partial \bar{p}^{(s)} }
\right] \Psi^{(c)}\ =\  c\ \Psi^{(c)}\,,
\end{eqnarray}
which are  quantum counterparts of constraints \p{M-constr-def-n}-\p{const-tw-4-n}
in the representation \p{realiz-var}.
In equation
\p{const-tw-4-qq} we have introduced the constant
 parameter $c$ related to the ambiguity of operator ordering
(it is an analog of the vacuum energy in the model of quantum
oscillator).
We emphasize that the equations
(\ref{M-constr-def-q}), (\ref{constr-tw-q}), (\ref{const-tw-4-qq})
are obtained in the framework of the Dirac canonical quantization
scheme of gauge systems where the constraints are imposed on the
wave functions depending on true physical degrees of freedom. As a
result we get the equations of motion describing the mass shell of
the model under consideration.

Equations \p{M-constr-def-q} and \p{constr-tw-q} can be solved
explicitly and solution is
\begin{equation}
\label{wf-sol}
\Psi^{(c)} \ = \ \delta\left(p^{(s)}\bar{p}^{(s)}-1\right)\,e^{\displaystyle -i(p^{(t)}+\bar{p}^{(t)})}\,
\tilde\Psi^{(c)}\big( p^{(z)}_{\alpha},\, \bar{p}^{(z)}_{\dot\alpha};\,
({\displaystyle p^{(s)}/\bar{p}^{(s)}})^{1/2}\big)\,,
\end{equation}
where $\tilde\Psi^{(c)}\big( p^{(z)}_{\alpha},\, \bar{p}^{(z)}_{\dot\alpha};\,
({\displaystyle p^{(s)}/\bar{p}^{(s)}})^{1/2}\big)$
is the function of the
complex coordinates $ p^{(z)},\bar{p}^{(z)}\! \in \, \stackrel{0}{\mathbb{C}}\!{}^2 = \mathbb{C}^2\backslash(0,0)$
and coordinate $({\displaystyle p^{(s)}/\bar{p}^{(s)}})^{1/2}= e^{i\varphi}$
on the unit circle: $p^{(s)} \cdot \bar{p}^{(s)} =1$,
i.e. $p^{(s)}= 
\exp(i\varphi)$ and
${\displaystyle p^{(s)}/\bar{p}^{(s)}} =\exp(2 i\varphi)$.
In general, the function $\tilde\Psi$ which depends  on
 $({\displaystyle p^{(s)}/\bar{p}^{(s)}})^{1/2}  = e^{i\varphi}$
 can be expanded into the Fourier series:
\begin{equation}
\label{Fourier-series}
\tilde\Psi^{(c)}\big( p^{(z)},\, \bar{p}^{(z)}; \,
({\displaystyle p^{(s)}/\bar{p}^{(s)}})^{1/2}\big) \ = \
\sum\limits_{k=-\infty}^{\infty} \left({\displaystyle \frac{p^{(s)}}{\bar{p}^{(s)}}}\right)^{-k/2}\,
\tilde\psi^{(c+k)}\big( p^{(z)},\, \bar{p}^{(z)}\big)\,.
\end{equation}
Therefore, solution \p{wf-sol} to the wave function
can be written in the form
\begin{equation}
\label{wf-sol-1}
\Psi^{(c)} \ = \  \delta\left(p^{(s)}\cdot \bar{p}^{(s)}-1\right)\; e^{\displaystyle -i(p^{(t)}+\bar{p}^{(t)})}
\,\sum\limits_{k=-\infty}^{\infty} \left({\displaystyle \frac{p^{(s)}}{\bar{p}^{(s)}}}\right)^{-k/2}\,
\tilde\psi^{(c+k)}\big( p^{(z)},\, \bar{p}^{(z)}\big) \,.
\end{equation}
Due to the last constraint \p{const-tw-4-qq} the coefficients
$\tilde\psi^{(c+k)}( p_z,\, \bar{p}_z)$ in
expansion \p{wf-sol-1} satisfy the equations
\begin{equation}
\label{const-tw-4-q}
\frac{1}{2}\,\left(p^{(z)}_{\alpha}\frac{\partial}{\partial p^{(z)}_{\alpha}} \ -\
\bar{p}^{(z)}_{\dot\alpha}\frac{\partial}{\partial \bar{p}^{(z)}_{\dot \alpha}}\right)
\tilde\psi^{(c+k)}\big( p^{(z)},\bar{p}^{(z)}\big)\ =\  \big(c+k\big) \, \tilde\psi^{(c+k)}\big( p^{(z)},\bar p^{(z)}\big)\,.
\end{equation}

Expression  \p{wf-sol-1} contains the delta-function $\delta(p^{(s)}\cdot \bar{p}^{(s)}-1)$.
Therefore, in formula \p{wf-sol-1} we can make replacements
\begin{equation}
\label{repl-delta}
\begin{array}{rcl}
\delta\left(p^{(s)}\cdot \bar{p}^{(s)}-1\right)\, p^{(s)}/\bar{p}^{(s)} & \to &
 \delta\left(p^{(s)}\cdot \bar{p}^{(s)}-1\right)\, (p^{(s)})^2\,, \\ [7pt]
\delta\left(p^{(s)}\cdot \bar{p}^{(s)}-1\right)\, \bar{p}^{(s)}/p^{(s)} & \to &
\delta\left(p^{(s)}\cdot \bar{p}_s-1\right)\, (\bar{p}^{(s)})^2\,,
\end{array}
\end{equation}
and then write the wave function   \p{wf-sol-1} as
\begin{eqnarray}
\nonumber
\Psi^{(c)} & = & \delta\left(p^{(s)}\cdot \bar{p}^{(s)}-1\right)\,e^{\displaystyle -i(p^{(t)}+\bar{p}^{(t)})}
\left(\!\tilde\psi^{(c)}\big( p^{(z)},\bar p^{(z)}\big) +
\sum\limits_{k=1}^{\infty} \big(\bar p^{(s)}\big)^{k}\,
\tilde\psi^{(c+k)}\big( p^{(z)},\bar p^{(z)}\big) \right. \\ [8pt]
&& \left. \hspace{7.5cm} + \
\sum\limits_{k=1}^{\infty} \big(p^{(s)}\big)^{k}\,
\tilde\psi^{(c-k)}\big( p^{(z)},\bar p^{(z)}\big)
\!\right).  \label{wf-sol-2}
\end{eqnarray}

Now we use relations  \eqref{def-new-var-z} and \eqref{def-new-var-bz}
to restore the dependence of the wave function \eqref{wf-sol-1} on the twistor variables.
As a result, we obtain the twistor wave function in the form (we leave the
same notation for functions as in \eqref{wf-sol-1})
\begin{equation}
\label{wf-tw}
\Psi^{(c)}( \pi,\bar \pi;\rho,\bar\rho) \ = \
\delta\left((\pi\rho)(\bar\rho\bar\pi)-M^2\right)\,
e^{\displaystyle -i\left(\frac{\rho_{1}}{\pi_{1}}+\frac{\bar\rho_{1}}{\bar\pi_{1}}\right)}\,
\tilde\Psi^{(c)}( \pi,\bar \pi;\rho,\bar\rho)\,,
\end{equation}
where
$\tilde\Psi^{(c)}( \pi,\bar \pi;\rho,\bar\rho)$ has the form
\begin{equation}
\label{wf-sol-tw-2}
\tilde\Psi^{(c)}( \pi,\bar \pi;\rho,\bar\rho) \ = \ \tilde\psi^{(c)}( \pi,\bar \pi) +
\sum\limits_{k=1}^{\infty} (\bar\rho\bar\pi)^{k}\,
\tilde\psi^{(c+k)}(\pi,\bar \pi) +
\sum\limits_{k=1}^{\infty} (\pi\rho)^{k}\,
\tilde\psi^{(c-k)}(\pi,\bar \pi) \; ,
\end{equation}
and the coefficients
$\tilde\psi^{(c\pm k)}\big( \pi,\bar \pi\big) : = M^{-|c\pm k|}
\, \tilde\psi^{(c\pm k)}\big( p^{(z)},\bar p^{(z)}\big)$
(which are the functions on the two-dimensional complex plane
$\stackrel{0}{\mathbb{C}}\!{}^2=\mathbb{C}^2\backslash(0,0)$ with the coordinates
$\pi_{\alpha}$) are subjected to the condition
\begin{equation}
\label{const-tw-4-q1}
\frac{1}{2}\,\left(\pi_{\alpha}\frac{\partial}{\partial \pi_{\alpha}} \ -\
\bar \pi_{\dot\alpha}\frac{\partial}{\partial \bar \pi_{\dot \alpha}}\right)
\tilde\psi^{(c\pm k)}( \pi,\bar \pi)\ =\  \big(c\pm k\big)\,
\tilde\psi^{(c\pm k)}( \pi,\bar \pi)\,.
\end{equation}
In \eqref{wf-tw} we have used the concise notation:
$(\pi\rho):=\pi^{\beta}\rho_{\beta}$,
$(\bar\rho\bar\pi):=\bar\rho_{\dot\beta}\bar\pi^{\dot\beta}$.

Under the shift of variables $\rho_{\alpha}$ and $\bar\rho_{\dot\alpha}$,
the twistor wave function, obtained in \eqref{wf-tw},
is transformed in the following way:
\begin{equation}
\label{sym-wf-tw}
\Psi^{(c)} ( \pi_{\alpha},\bar \pi_{\dot\alpha};\rho_{\alpha}+\kappa\,\pi_{\alpha},
\bar\rho_{\dot\alpha}+\bar\kappa\,\bar\pi_{\dot\alpha}) \ = \
e^{\displaystyle -i\left(\kappa+\bar\kappa\right)}\,
\Psi^{(c)} ( \pi_{\alpha},\bar \pi_{\dot\alpha};\rho_{\alpha},\bar\rho_{\dot\alpha})\,,
\;\;\;\; \forall \kappa \in \mathbb{C} \; ,
\end{equation}
and, therefore, the wave function \eqref{wf-tw}
satisfies the equations
\begin{equation}
\label{equ-wf-tw}
i\,\pi_{\alpha}\,\frac{\partial}{\partial\rho_{\alpha}}\,\Psi^{(c)} =  \Psi^{(c)} \,,\qquad
i\,\bar\pi_{\dot\alpha}\,\frac{\partial}{\partial\bar\rho_{\dot\alpha}}\,\Psi^{(c)}
=  \Psi^{(c)} \,,
\end{equation}
which are quantum counterparts of constraints  \eqref{constr-tw}.
Also, the twistor wave function  \eqref{wf-tw} obeys the condition
\begin{equation}
\label{equ-wf-tw-2}
\left(\pi_{\alpha}\frac{\partial}{\partial \pi_{\alpha}} \ -\
\bar \pi_{\dot\alpha}\frac{\partial}{\partial \bar \pi_{\dot \alpha}} \ + \
\rho_{\alpha}\frac{\partial}{\partial \rho_{\alpha}} \ -\
\bar \rho_{\dot\alpha}\frac{\partial}{\partial \bar \rho_{\dot \alpha}}\right)\,
\Psi^{(c)} \ =\   2c \,  \Psi^{(c)}\,,
\end{equation}
which is a quantum counterpart of constraint   \eqref{const-tw-4}.
Equation \eqref{equ-wf-tw-2} is equivalent to the homogeneity
condition
\begin{equation}
\label{equ-wf-tw-3}
\Psi^{(c)} ( \mathrm{e}^{i\gamma}\pi_{\alpha},\mathrm{e}^{-i\gamma}\bar \pi_{\dot\alpha};
\mathrm{e}^{i\gamma}\rho_{\alpha},\mathrm{e}^{-i\gamma}\bar\rho_{\dot\alpha}) \ =\
\mathrm{e}^{2ic\gamma}\Psi^{(c)} ( \pi_{\alpha},\bar \pi_{\dot\alpha};\rho_{\alpha},\bar\rho_{\dot\alpha})
\end{equation}
of the twistor field $\Psi^{(c)}$ with respect
to $\mathrm{U}(1)$ transformations with the constant phase parameter $\gamma$,
where the constant $c$ plays the role of the $\mathrm{U}(1)$  charge.
The uniqueness condition of the twistor field requires integer or half-integer values for the constant $c$:
\begin{equation}
\label{integer-c}
2c \ \in \ \mathbb{Z} \,.
\end{equation}

Now we check that the twistor wave function \eqref{wf-tw} describes the massless particle of the infinite (continuous) spin.
To do this, we use
the twistor realization of the Poincar\'{e} algebra generators
\begin{equation}
\label{P-op}
\mathbb{P}_{\alpha\dot\beta}= \pi_{\alpha} \bar\pi_{\dot\beta}  \,,\qquad
\mathbb{M}_{\alpha\beta} = \pi_{(\alpha}\,\frac{\partial}{\partial\pi^{\beta)}}+
\rho_{(\alpha}\,\frac{\partial}{\partial\rho^{\beta)}}\,,\qquad
\bar{\mathbb{M}}_{\dot\alpha\dot\beta} = \bar\pi_{(\dot\alpha}\,\frac{\partial}{\partial\bar\pi^{\dot\beta)}}+
\bar\rho_{(\dot\alpha}\,\frac{\partial}{\partial\bar\rho^{\dot\beta)}}\, ,
\end{equation}
in terms of which the Pauli-Luba\'{n}ski operator
\begin{equation}
\label{PL-sp-op}
\mathbb{W}_{\alpha\dot\gamma}=
\bar {\mathbb{M}}_{\dot\gamma\dot\beta}\mathbb{P}^{\dot\beta}_{\alpha} - \mathbb{M}_{\alpha\beta}\mathbb{P}^{\beta}_{\dot\gamma}  \; ,
\end{equation}
takes the form (see \cite{BFIR})
\begin{equation}
\label{PL-tw-op}
\mathbb{W}_{\alpha\dot\gamma}=
\pi_{\alpha}\bar\pi_{\dot\gamma}\,\mathbf{\Lambda}
+\frac{1}{2}\left[\pi_{\alpha}\bar\rho_{\dot\gamma} \Big(\bar\pi_{\dot\beta}\,\frac{\partial}{\partial\bar\rho_{\dot\beta}}\Big)
-\rho_{\alpha}\bar\pi_{\dot\gamma} \Big(\pi_{\beta}\,\frac{\partial}{\partial\rho_{\beta}}\Big)\right]
+\frac{1}{2}\left[(\bar\rho\bar\pi)\,\pi_{\alpha}\,\frac{\partial}{\partial\bar\rho^{\dot\gamma}}- (\pi\rho)\,\bar\pi_{\dot\gamma}\,\frac{\partial}{\partial\rho^{\alpha}}\right]\,,
\end{equation}
where
\begin{equation}
\label{hel-tw-op}
\mathbf{\Lambda}=-\frac{1}{2}\left(\pi_{\beta}\,\frac{\partial}{\partial\pi_{\beta}}-
\bar\pi_{\dot\beta}\,\frac{\partial}{\partial\bar\pi_{\dot\beta}}\right)\,.
\end{equation}
Direct calculations show that
\begin{equation}
\label{act-W-part}
\mathbb{W}_{\alpha\dot\gamma} \,\pi^{\beta}\rho_{\beta}=\mathbb{W}_{\alpha\dot\gamma} \,\bar\rho_{\dot\beta}\bar\pi^{\dot\beta}=0\,,\qquad
\mathbb{W}_{\alpha\dot\gamma} \left(\frac{\rho_{1}}{\pi_{1}}+\frac{\bar\rho_{1}}{\bar\pi_{1}}\right)=
(\pi\rho)\,\frac{\epsilon_{\alpha 1}\bar\pi_{\dot\gamma}}{\pi_1} \ - \
(\bar\rho\bar\pi)\,\frac{\epsilon_{\dot\gamma 1} \pi_{\alpha}}{\bar\pi_1}\,.
\end{equation}
By using \eqref{act-W-part} one can find
the action of the Pauli-Luba\'{n}ski operator  \eqref{PL-tw-op} on
the twistorial wave function  \eqref{wf-tw}:
\begin{equation}
\label{ct-W-wf-tw}
\mathbb{W}_{\alpha\dot\gamma} \Psi^{(c)}   =
\delta\left((\pi\rho)(\bar\rho\bar\pi)-M^2\right)\,
e^{\displaystyle -i\left(\frac{\rho_{1}}{\pi_{1}}+\frac{\bar\rho_{1}}{\bar\pi_{1}}\right)}
\,D_{\alpha\dot\gamma}\,\tilde\Psi^{(c)} \,,
\end{equation}
where the operator $D_{\alpha\dot\gamma}$, acting on the reduced twistor field $\tilde\Psi^{(c)}$, takes the form
\begin{equation}
\label{ct-D-wf-tw}
D_{\alpha\dot\gamma} \ :=
\ \pi_{\alpha}\bar\pi_{\dot\gamma}\,\mathbf{\Lambda}
+i\left((\bar\rho\bar\pi)\,\frac{\epsilon_{\dot\gamma 1} \pi_{\alpha}}{\bar\pi_1}
-(\pi\rho)\,\frac{\epsilon_{\alpha 1}\bar\pi_{\dot\gamma}}{\pi_1}\right)
\,.
\end{equation}
Acting on both sides
of \eqref{ct-W-wf-tw} by the operator $\mathbb{W}^{\alpha\dot\gamma}$ and
taking into account the identity
$D^{\alpha\dot\gamma}D_{\alpha\dot\gamma}\,\tilde\Psi=-2M^2\,\tilde\Psi$, we obtain that
\begin{equation}
\label{ct-W-wf-tw1}
\mathbb{W}^{\alpha\dot\gamma}\mathbb{W}_{\alpha\dot\gamma}\, \Psi^{(c)} \ = \
-2M^2\,\Psi^{(c)} \ = \ -\mu^2\,\Psi^{(c)}\,.
\end{equation}
So the twistor field  \eqref{wf-tw}
indeed describes a massless particle of the infinite spin.

The states with fixed helicities are the eigenvectors of the helicity operator $\mathbb{\Lambda}$ which is defined
as a projection of the total angular momentum $\vec{\mathbb{J}\,\,}$ in the direction of motion with the momentum operator $\mathbb{P}_m=(\mathbb{P}_0,\vec{\mathbb{P}})$: $\mathbb{\Lambda}=\vec{\mathbb{J}\,\,}\vec{\mathbb{P}}/\mathbb{P}_0$.
This operator can be rewritten in terms of the Pauli-Luba\'{n}ski pseudovector \eqref{PL-tw-op} in the form
\begin{equation}
\label{hel-exp}
\mathbb{\Lambda} \ = \ \frac{\mathbb{W}_{0}}{\mathbb{P}_{0}} \ = \
\frac{\mathbb{W}_{\alpha\dot\gamma}\tilde\sigma_0^{\dot\gamma\alpha}}{\pi_{\beta}\bar\pi_{\dot\beta}\tilde\sigma_0^{\dot\beta\beta}} \ = \
\frac{\sum\limits_{\alpha=\dot\gamma} \mathbb{W}_{\alpha\dot\gamma}}{\sum\limits_{\beta=\dot\beta} \pi_{\beta}\bar\pi_{\dot\beta}} \,.
\end{equation}
As we see from \eqref{ct-W-wf-tw} the representation \eqref{wf-tw} of the twistor field is unsuitable for helicity expansion
of the continuous spin wave function.

{}For finding helicity expansion of the twistor wave function  \eqref{wf-tw} we represent expression \eqref{wf-tw}
in the following form:\,\footnote{Here we take into account the  equality
$$
\frac{\rho_{1}}{\pi_{1}}+\frac{\bar\rho_{1}}{\bar\pi_{1}}+
\frac{(\bar\rho \bar\pi )\, \pi_{1}\pi_{2}+ (\pi \rho)\,\bar\pi_{1}\bar\pi_{2}}{\pi_{1}\bar\pi_{1}\sum\limits_{\beta=\dot\beta} \pi_{\beta}\bar\pi_{\dot\beta}} \ = \
\frac{\sum\limits_{\alpha=\dot\alpha} (\pi_{\alpha}\bar\rho_{\dot\alpha} + \rho_{\alpha}\bar\pi_{\dot\alpha})}
{\sum\limits_{\beta=\dot\beta} \pi_{\beta}\bar\pi_{\dot\beta}}\,.
$$
}
\begin{equation}
\label{wf-tw-hel}
\Psi^{(c)} (\pi,\bar \pi;\rho,\bar\rho)^{(c)} \ = \
\delta\left((\pi\rho)(\bar\rho\bar\pi)-M^2\right)\;
\exp \Bigl( {\displaystyle \frac{-i(\pi_{\gamma}\bar\rho_{\dot\gamma}+
\rho_{\gamma}\bar\pi_{\dot\gamma})\tilde\sigma_0^{\dot\gamma\gamma}}
{(\pi_{\delta}\bar\pi_{\dot\delta}\tilde\sigma_0^{\dot\delta\delta})}}\,
 \Bigr) \; \hat\Psi^{(c)} (\pi,\bar \pi;\rho,\bar\rho)\, ,
\end{equation}
where
\begin{equation}
\label{Psi-Psi}
\hat\Psi^{(c)}(\pi,\bar \pi;\rho,\bar\rho) \ = \
\exp \Bigl( {\displaystyle i\, \frac{(\bar\rho \bar\pi )\, \pi_{1}\pi_{2}+ (\pi\rho)\,\bar\pi_{1}\bar\pi_{2}}{\pi_{1}\bar\pi_{1}
(\pi_{\beta}\bar\pi_{\dot\beta}\tilde\sigma_0^{\dot\beta\beta})}} \Bigr)
 \; \tilde\Psi^{(c)}(\pi,\bar \pi;\rho,\bar\rho)\,,
\end{equation}
and the function $\tilde\Psi^{(c)}( \pi,\bar \pi;\rho,\bar\rho)$
is defined in \p{wf-sol-tw-2}.
In fact  the exponent in \eqref{wf-tw-hel} is
expressed in terms of zero components of four-vectors \eqref{p-resol} and  \eqref{w-resol}.
So formula \eqref{wf-tw-hel} takes the form
\begin{equation}
\label{wf-tw-hel1}
\Psi^{(c)} ( \pi,\bar \pi;\rho,\bar\rho) \ = \
\delta\left((\pi\rho)(\bar\rho\bar\pi)-M^2\right)\,
e^{\displaystyle -iq_0/p_0}\,
\hat\Psi^{(c)} (\pi,\bar \pi;\rho,\bar\rho)\,,
\end{equation}
in which the quantities $p_0$ and $q_0$ have the
generalized Cartan-Penrose representations  \eqref{p-resol} and  \eqref{w-resol}.

Expression of $\hat\Psi^{(c)} (\pi,\bar \pi;\rho,\bar\rho)$, defined in \eqref{Psi-Psi},
contains the spinors $\rho_\alpha$, $\bar\rho_{\dot\alpha}$ only in the
contractions $(\pi\rho)$, $(\bar\rho \bar\pi )$.
But, expanding $\exp\Big({ i\, \frac{(\bar\rho \bar\pi )\, \pi_{1}\pi_{2}+ (\pi\rho)\,\bar\pi_{1}\bar\pi_{2}}{\pi_{1}\bar\pi_{1}
(\pi_{\beta}\bar\pi_{\dot\beta}\tilde\sigma_0^{\dot\beta\beta})}}\Big)$
in  \eqref{Psi-Psi} as the Fourier series and using $\delta\left((\pi\rho)(\bar\rho\bar\pi)-M^2\right)$,
we can represent $\hat\Psi^{(c)} (\pi,\bar \pi;\rho,\bar\rho)$ in the form
\begin{equation}
\label{wf-sol-tw-4}
\hat\Psi^{(c)}( \pi,\bar \pi;\rho,\bar\rho) \ = \ \psi^{(c)}( \pi,\bar \pi) +
\sum\limits_{k=1}^{\infty} (\bar\rho\bar\pi)^{k}\,
\psi^{(c+k)}(\pi,\bar \pi) +
\sum\limits_{k=1}^{\infty} (\pi\rho)^{k}\,
\psi^{(c-k)}(\pi,\bar \pi)\,,
\end{equation}
similar to \p{wf-sol-tw-2} for $\tilde\Psi( \pi,\bar \pi;\rho,\bar\rho)$.
The fields $\psi^{(c\pm k)}( \pi,\bar \pi)$ are subjected to the conditions
\begin{equation}
\label{const-tw-4-q1-n}
\frac{1}{2}\,\left(\pi_{\alpha}\frac{\partial}{\partial \pi_{\alpha}} \ -\
\bar \pi_{\dot\alpha}\frac{\partial}{\partial \bar \pi_{\dot \alpha}}\right)
\psi^{(c\pm k)}( \pi,\bar \pi)\ =\  \big(c\pm k\big)\,  \psi^{(c\pm k)}( \pi,\bar \pi)\,,
\end{equation}
similar to \eqref{const-tw-4-q1}.

The representation \eqref{wf-tw-hel1} for the twistorial field of the infinite spin particle is convenient
to clarify its helicity content. Considering the relation
\begin{eqnarray}
\label{act-W-part1}
\mathbb{W}_{\alpha\dot\alpha} \,\frac{q_{0}}{p_{0}}&=&
\frac{1}{2}\,\bigg[\pi_{\alpha}\bar\rho_{\dot\alpha} - \rho_{\alpha}\bar\pi_{\dot\alpha} -
\frac{\pi_{\alpha}\bar\pi_{\dot\alpha}}{\sum\limits_{\delta=\dot\delta}\pi_{\delta}\bar\pi_{\dot\delta}} \,
\sum\limits_{\beta=\dot\beta} (\pi_{\beta}\bar\rho_{\dot\beta} - \rho_{\beta}\bar\pi_{\dot\beta}) \bigg]\\
&&-\, \frac{1}{2\sum\limits_{\delta=\dot\delta}\pi_{\delta}\bar\pi_{\dot\delta}} \,
\bigg[(\bar\rho\bar\pi)
\sum\limits_{\beta=\dot\beta} (\pi_{\alpha} \epsilon_{\dot\alpha\dot\beta}\pi_{\beta}) -
(\pi\rho) \sum\limits_{\beta=\dot\beta} (\bar\pi_{\dot\alpha}\epsilon_{\alpha\beta} \bar\pi_{\dot\beta})\bigg]\,, \nonumber
\end{eqnarray}
which holds for the operator \eqref{PL-tw-op} and using the generalized
Cartan-Penrose representations  \eqref{p-resol}, \eqref{w-resol} for $p_0$, $q_0$,
we obtain that the action of the Pauli-Luba\'{n}ski pseudovector
on the field \eqref{wf-tw-hel1}
gives
\begin{equation}
\label{ct-W-wf-tw-h}
\! \mathbb{W}_{\alpha\dot\alpha} \Psi^{(c)}\!   =
\delta\!\left((\pi\rho)(\bar\rho\bar\pi)-M^2\right)
e^{\displaystyle -iq_0/p_0}\!
\left( \hat D_{\alpha\dot\alpha}\psi^{(c)} +
\sum\limits_{k=1}^{\infty} (\bar\rho\bar\pi)^{k}
\hat D_{\alpha\dot\alpha}\psi^{(c+k)} +
\sum\limits_{k=1}^{\infty} (\pi\rho)^{k}\,
\hat D_{\alpha\dot\alpha}\psi^{(c-k)}\right) ,
\end{equation}
where
\begin{eqnarray}
\label{D-new}
\hat D_{\alpha\dot\alpha}&=&{ \displaystyle \pi_{\alpha}\bar\pi_{\dot\alpha}\,\mathbf{\Lambda}
- \frac{i}{2}\,\big(\pi_{\alpha}\bar\rho_{\dot\alpha}  -  \rho_{\alpha}\bar\pi_{\dot\alpha}\big)}
\\ [7pt]
&& { \displaystyle  + \
\frac{i\pi_{\alpha}\bar\pi_{\dot\alpha}}{2\sum\limits_{\delta=\dot\delta}\pi_{\delta}\bar\pi_{\dot\delta}} \,
\sum\limits_{\beta=\dot\beta} (\pi_{\beta}\bar\rho_{\dot\beta} - \rho_{\beta}\bar\pi_{\dot\beta})   +
\frac{i}{2\sum\limits_{\delta=\dot\delta}\pi_{\delta}\bar\pi_{\dot\delta}} \,
\bigg[(\bar\rho\bar\pi)
\sum\limits_{\beta=\dot\beta} (\pi_{\alpha} \epsilon_{\dot\alpha\dot\beta}\pi_{\beta}) -
(\pi\rho) \sum\limits_{\beta=\dot\beta} (\bar\pi_{\dot\alpha}\epsilon_{\alpha\beta} \bar\pi_{\dot\beta})\bigg]
\,.}\nonumber
\end{eqnarray}
In contrast to quantity \eqref{ct-D-wf-tw}, the operator  \eqref{D-new} satisfies the property
$$
\sum\limits_{\alpha=\dot\alpha}\hat D_{\alpha\dot\alpha}=\sum\limits_{\alpha=\dot\alpha}\pi_{\alpha}
\bar\pi_{\dot\alpha}\,\mathbf{\Lambda}=
\mathbb{P}_{0}\mathbf{\Lambda}\,.
$$
As a result, the helicity operator \eqref{hel-exp} acts on
the twistorial field in the following way:
\begin{eqnarray}
\label{hel-act-wf}
\mathbb{\Lambda} \, \Psi^{(c)} &=& \frac{\mathbb{W}_{0}}{\mathbb{P}_{0}}\, \Psi^{(c)} \ = \
\frac{\sum\limits_{\alpha=\dot\alpha}\mathbb{W}_{\alpha\dot\alpha}}{
\sum\limits_{\beta=\dot\beta}\pi_{\beta}\bar\pi_{\dot\beta}} \, \Psi^{(c)}  \\ [6pt]
&=& \delta\left((\pi\rho)(\bar\rho\bar\pi)-M^2\right)
e^{\displaystyle -iq_0/p_0}
\, \left( \mathbf{\Lambda}\psi^{(c)} +
\sum\limits_{k=1}^{\infty} (\bar\rho\bar\pi)^{k}
\mathbf{\Lambda}\psi^{(c+k)} +
\sum\limits_{k=1}^{\infty} (\pi\rho)^{k}\,
\mathbf{\Lambda}\psi^{(c-k)}\right) \,.
\nonumber
\end{eqnarray}
As we see from this expression, the action of the helicity operator $\mathbb{\Lambda}$ is
defined by the action of the operator $\mathbf{\Lambda}$ on the functions
$\psi^{(c+k)}(\pi,\bar \pi)$ (where $k \in \mathbb{Z}$) which
are parameterized by two dimensonal complex variable $\pi_{\alpha}$.
Due to the conditions \eqref{const-tw-4-q1-n}, the  fields $\psi^{(c+k)}( \pi,\bar \pi)$ are
the eigenvectors of the  helicity operator \eqref{hel-tw-op}:
\begin{equation}
\label{const-tw-4-q1-s}
\mathbf{\Lambda}\,
\psi^{(c+k)}( \pi,\bar \pi)\ =\  - \big(c+k\big)\,
\psi^{(c+k)}( \pi,\bar \pi)\, .
\end{equation}
Thus, in view of this equation and \eqref{hel-act-wf}, the
twistorial wave function \eqref{wf-tw-hel1} of the infinite spin
particle describes an infinite number of massless states
$\psi^{(c+k)}$ whose helicities are equal to integer (for integer
$c$) or half-integer (for half-integer $c$) values and these
helicities run from $-\infty$ to $+\infty$.

Emphasize that the helicity  content of the twistor field
$\Psi^{(c)}$ is the same for all integer values of the
$\mathrm{U}(1)$ charge $c$. The distinction is only in the shift of
the infinite tower of states (with all possible integer helicities)
in $k$ by integer difference of the values of $c$.
 The same situation occurs for half-integer values
 of $c$, when the tower of
states with half-integer helicities is also the same for various $c
\in (\mathbb{Z} + 1/2)$. The choice of number $c$, which takes a
fixed value, determines only a specific way of describing the same
infinite spin representations for all integer (or half-integer)
spins. In fact, there are only two independent values for $c$,
namely $c=0$ and $c=-\frac{1}{2}.$ In this sense the sets
$\{c+k\}$ with integer $k$ and $\{c\}$ are equivalent and we can use
the any representative of the set $\{c+k\}$.

Recall that
the twistorial wave function \eqref{wf-tw-hel1} is complex and
therefore all component fields in the expansion \eqref{wf-sol-tw-4} are also complex.
Then, in the CPT-invariant theory we must consider together with the field $\Psi^{(c)}$
its complex conjugated field  $(\Psi^{(c)})^*$ which
(due to the condition \eqref{equ-wf-tw-2}) has the opposite charge
\begin{equation}
\label{c-c}
(\Psi^{(c)})^*:=\bar\Psi^{(-c)} \; .
\end{equation}
To describe the bosonic infinite spin representation
related to all integer helicities, we put
\begin{equation}
\label{q-0}
c=0
\end{equation}
and consider the twistorial field
\begin{equation}
\label{wf-tw-0}
\Psi^{(0)} ( \pi,\bar \pi;\rho,\bar\rho) \; .
\end{equation}
The complex conjugate field $\bar\Psi^{(0)}$ has also zero charge.
Similarly, to describe the infinite spin representation related to
half-integer helicities we take for $c$ the value
\begin{equation}
\label{q-12}
c=- \,\frac12 \, .
\end{equation}
The corresponding wave function
\begin{equation}
\label{wf-tw-12}
\Psi^{(-1/2)} ( \pi,\bar \pi;\rho,\bar\rho)
\end{equation}
contains in its expansion  \eqref{wf-sol-tw-4} all half-integer helicities.
The complex conjugate field
\begin{equation}
\label{wf-tw-12c}
\bar\Psi^{(+1/2)} ( \pi,\bar \pi;\rho,\bar\rho)
\end{equation}
possesses the charge $c=+1/2$.

To conclude this section, we stress that the choice of the integer
or half-integer values of the $\mathrm{U}(1)$-charge\footnote{The
interpretation of the parameter $c$ as $\mathrm{U}(1)$-charge
follows from equations \p{equ-wf-tw-2} and \p{equ-wf-tw-3}.}  $c$
can be treated as choice of different boundary conditions for
equations of motion, by analogy with the choice of the Ramond or
Nevew-Schwarz sectors in the superstring theories (see, e.g.,
\cite{GSW}).

\setcounter{equation}{0}
\section{Twistor transform for infinite spin fields}

\quad\, In \cite{BFIR}, performing a quantization of the twistor model with
special gauge fixing, we obtained
space-time fields which depended on the coordinates $x^m$ of position four-vector
and obeyed the Wigner-Bargmann equations \cite{Wigner47,BargWigner} following
from Lagrangian (\ref{L-sp}).
Now we show that our twistorial model reproduces the formulation
of the infinite spin field model developed in \cite{BuchKrTak}.

The link of our twistor field with the fields introduced in \cite{BuchKrTak}
follows from the explicit solutions
\eqref{wf-tw-hel1} and  \eqref{wf-sol-tw-4} obtained for the twistor wave functions.
For further convenience we introduce the dimensionless spinor
\begin{equation}
\label{spinor-new}
\xi_{\alpha}:=M^{-1/2}\rho_{\alpha}\,,\qquad
\bar \xi_{\dot\alpha}:=M^{-1/2} \bar \rho_{\dot \alpha}\,.
\end{equation}
Then, the twistor wave function \eqref{wf-tw-0} of infinite integer-spin particle
(see \eqref{wf-tw-hel1} and  \eqref{wf-sol-tw-4}) takes the form
\begin{eqnarray}
\label{wf-tw-hel-0}
\Psi^{(0)}( \pi,\bar \pi;\xi,\bar\xi) &=&
\delta\left((\pi\xi)(\bar\xi\bar\pi)-M\right)\,
e^{\displaystyle -iq_0/p_0}\,
\hat\Psi^{(0)} (\pi,\bar \pi;\xi,\bar\xi)\,,
\\ [6pt]
\nonumber
&& \hat\Psi^{(0)} \ = \ \psi^{(0)}( \pi,\bar \pi) +
\sum\limits_{k=1}^{\infty} (\bar\xi\bar\pi)^{k}\,
\psi^{(k)}(\pi,\bar \pi) +
\sum\limits_{k=1}^{\infty} (\pi\xi)^{k}\,
\psi^{(-k)}(\pi,\bar \pi)\,.
\end{eqnarray}
In the expansion of $\hat\Psi^{(0)}$
all components $\psi^{(k)}( \pi,\bar \pi)$ (here $k\in \mathbb{Z}$)
are complex functions (fields), in general.
The wave function of the infinite half-integer spin particle \eqref{wf-tw-12} is
\begin{eqnarray}
\label{wf-tw-hel-12}
\Psi^{(-\frac12)}( \pi,\bar \pi;\xi,\bar\xi) &=&
\delta\left((\pi\xi)(\bar\xi\bar\pi)-M\right)\,
e^{\displaystyle -iq_0/p_0}\,
\hat\Psi^{(-\frac12)} (\pi,\bar \pi;\xi,\bar\xi)\,,
\\ [6pt]
\nonumber
&& \hat\Psi^{(-\frac12)} \ = \ \psi^{(-\frac12)}( \pi,\bar \pi) +
\sum\limits_{k=1}^{\infty} (\bar\xi\bar\pi)^{k}\,
\psi^{(-\frac12+k)}(\pi,\bar \pi) +
\sum\limits_{k=1}^{\infty} (\pi\xi)^{k}\,
\psi^{(-\frac12-k)}(\pi,\bar \pi)\,.
\end{eqnarray}
The expansion of the complex conjugated wave function \eqref{wf-tw-12c} is
\begin{eqnarray}
\label{wf-tw-hel-12c}
\bar\Psi^{(+\frac12)}( \pi,\bar \pi;\xi,\bar\xi) &=&
\delta\left((\pi\xi)(\bar\xi\bar\pi)-M\right)\,
e^{\displaystyle iq_0/p_0}\,
\hat{\bar\Psi}^{(+\frac12)} (\pi,\bar \pi;\xi,\bar\xi)\,,
\\ [6pt]
\nonumber
&& \hat{\bar\Psi}^{(+\frac12)} \ = \ \bar\psi^{(\frac12)}( \pi,\bar \pi) +
\sum\limits_{k=1}^{\infty} (\bar\xi\bar\pi)^{k}\,
\bar\psi^{(\frac12+k)}(\pi,\bar \pi) +
\sum\limits_{k=1}^{\infty} (\pi\xi)^{k}\,
\bar\psi^{(\frac12-k)}(\pi,\bar \pi) \, ,
\end{eqnarray}
where
the components (fields)
$\bar\psi^{(r)}(\pi,\bar \pi)$ are complex conjugation of the
components (fields) $\psi^{(-r)}(\pi,\bar \pi)$
in the expansion of \eqref{wf-tw-hel-12}:
\begin{equation}
\label{cf-con-12}
\left(\psi^{(-\frac12+k)}\right)^*=\bar\psi^{(\frac12-k)}\; ,\qquad  k\in \mathbb{Z}\,.
\end{equation}

Recall that the quantities $p_0$ and $q_0$
in \eqref{wf-tw-hel-0}, \eqref{wf-tw-hel-12}, \eqref{wf-tw-hel-12c}
can be resolved
 by means of the generalized Cartan-Penrose representations  \eqref{p-resol} and  \eqref{w-resol},
 so we have
\begin{equation}
\label{w0-p0}
\frac{q_0}{p_0} \ = \
\frac{\sqrt{M}\sum\limits_{\alpha=\dot\alpha}(\pi_{\alpha}\bar\xi_{\dot\alpha} + \xi_{\alpha}\bar\pi_{\dot\alpha})}
{\sum\limits_{\beta=\dot\beta}\pi_{\beta}\bar\pi_{\dot\beta}} \,.
\end{equation}
In view of \eqref{const-tw-4-q1-n}
we assign the $\mathrm{U}(1)$-charges $\big(c\pm k\big)$ to the
the component fields $\psi^{(c\pm k)}(\pi,\bar \pi)$ in
\eqref{wf-tw-hel-0}, \eqref{wf-tw-hel-12}, \eqref{wf-tw-hel-12c}.
Note that in \eqref{wf-tw-hel-0}, \eqref{wf-tw-hel-12}, \eqref{wf-tw-hel-12c}
we use the same notation for the component fields $\psi^{(k)}(\pi,\bar \pi)$ as
in \eqref{wf-sol-tw-4} so as not to clutter the expressions by new notation.
 Of course,  the fields $\psi^{(c\pm k)}(\pi,\bar \pi)$
in \eqref{wf-tw-hel-0}, \eqref{wf-tw-hel-12}, \eqref{wf-tw-hel-12c}
equal the fields $\psi^{(c\pm k)}(\pi,\bar \pi)$ in
\eqref{wf-sol-tw-4}  up to multipliers $M^{k/2}$.

The $\delta$-function in \eqref{wf-tw-hel-0}, \eqref{wf-tw-hel-12}, \eqref{wf-tw-hel-12c} has the form
\begin{equation}
\label{delta-i}
\delta\left((\pi\xi)(\bar\xi\bar\pi)-M\right)=\delta\left(\xi^\alpha p_{\alpha\dot\alpha}\bar\xi^{\dot\alpha}-M\right)\,,
\end{equation}
if we take into account $p_{\alpha\dot\alpha}=\pi_{\alpha}\bar\pi_{\dot\alpha}$.
Such $\delta$-function is present in the definition of fields in  \cite{BuchKrTak}.
Moreover, the residual part
$e^{\displaystyle -iq_0/p_0}\,\hat\Psi^{(c)} (\pi,\bar \pi;\xi,\bar\xi)$ of the field
\eqref{wf-tw-hel-0}, \eqref{wf-tw-hel-12}, \eqref{wf-tw-hel-12c}
is a polynomial in the spinor variables $\xi_{\alpha}$, as it is in \cite{BuchKrTak}.
So we can interpret the wave functions
  \eqref{wf-tw-hel-0}, \eqref{wf-tw-hel-12}, \eqref{wf-tw-hel-12c}
as the fields in the momentum representation,
where momentum $p_{m}$ is defined by the spinors  $\pi_{\alpha}$
and $\bar\pi_{\dot\alpha}$.

\subsection{Integer spins}

\quad\,
We take $c=0$ for the value of the $\mathrm{U}(1)$-charge $c$ of
the twistor field $\Psi^{(c)}$ as in \eqref{q-0}.
The field $\Psi^{(0)}$ in this case is given in \eqref{wf-tw-hel-0}.
Consider the space-time wave function determined by means of the integral
Fourier transformation
\begin{equation}
\label{wf-st-tw}
\Phi(x;\xi,\bar\xi) \ = \ \int d^4 \pi \, e^{\displaystyle \,i p_{\alpha\dot\alpha} x^{\dot\alpha\alpha}}\,
\Psi^{(0)} (\pi,\bar\pi;\xi,\bar\xi)\ = \ \int d^4 \pi \, e^{\displaystyle \,i \pi_{\alpha}\bar\pi_{\dot\alpha} x^{\dot\alpha\alpha}}\,
\Psi^{(0)} (\pi,\bar\pi;\xi,\bar\xi)\,,
\end{equation}
where
we have used for the momentum
the representation $p_{\alpha\dot\alpha}=\pi_{\alpha}\bar\pi_{\dot\alpha}$
(see \eqref{p-resol}).
In the integral \eqref{wf-st-tw} we perform integration
over the two-dimensional complex space with the integration measure
$d^4 \pi := \frac18\, d\pi^{\alpha}\wedge d\pi_{\alpha}\wedge d\bar\pi^{\dot\alpha}\wedge d\bar\pi_{\dot\alpha} =
\frac12\, d\pi_{1}\wedge d\pi_{2}\wedge d\bar\pi_{\dot1}\wedge d\bar\pi_{\dot2}$.
This integration is in fact the integration over four-vector $p = (p_0,p_1,p_2,p_3)$
with measure $d^4p \; \delta(p^2)$ (see Appendix\,B) and integration over
common phase in $\pi_\alpha$ which is not presented in $p_{\alpha\dot\alpha}=\pi_{\alpha}\bar\pi_{\dot\alpha}$.

The field \eqref{wf-st-tw} satisfies four equations
\begin{eqnarray}
\label{KG-tw}
\partial^{\alpha\dot\alpha}\partial_{\alpha\dot\alpha}\,\Phi(x;\xi,\bar\xi) & = & 0 \,,
\\ [7pt]
\label{1eq-tw}
\left(i\xi^{\alpha}\partial_{\alpha\dot\alpha}\bar\xi^{\dot\alpha}+M\right)\Phi(x;\xi,\bar\xi) & = & 0 \,,
\\ [7pt]
\label{2eq-tw}
\left(i\frac{\partial}{\partial\xi_{\alpha}}\partial_{\alpha\dot\alpha}
\frac{\partial}{\partial\bar\xi_{\dot\alpha}}-M\right)\Phi(x;\xi,\bar\xi) & = & 0 \,,
\\ [7pt]
\label{3eq-tw}
\left(\xi_{\alpha}\frac{\partial}{\partial\xi_{\alpha}}-\bar\xi_{\dot\alpha}
\frac{\partial}{\partial\bar\xi_{\dot\alpha}}\right)\Phi(x;\xi,\bar\xi) & = & 0 \,.
\end{eqnarray}
The proof of these equations
is given in Appendix\,A. Equations \eqref{KG-tw}, \eqref{1eq-tw},
\eqref{2eq-tw}, \eqref{3eq-tw} underlie the definition
of the infinite spin fields proposed in \cite{BuchKrTak}.
The explicit form of the momentum wave function corresponding to the space-time field \eqref{wf-st-tw}
is given in Appendix\,B.

\subsection{Half-integer spins}

\quad\,
In this case, for the $\mathrm{U}(1)$-charge $c$ we choose the value
 $c=- 1/2$  (see \eqref{q-12}). The corresponding twistor field
 $\Psi^{(-1/2)}$ is written in \eqref{wf-tw-hel-12}.
Then we use the standard prescription of the twistorial
definition of space-time fields with nonvanishing helicities.
Namely, we have to insert the twistorial spinor $\pi_\alpha$ in the integrand in the integral Fourier
transformation  \eqref{wf-st-tw}.
Therefore, in the case $c=- 1/2$ the space-time field
for describing the half-integer spins has the form
\begin{equation}
\label{wf-st-tw-12}
\Phi_\alpha(x;\xi,\bar\xi) \ = \ \int d^4 \pi \, e^{\displaystyle \,i p_{\beta\dot\beta} x^{\dot\beta\beta}}\,\pi_\alpha\,
\Psi^{(-1/2)} (\pi,\bar\pi;\xi,\bar\xi)\ = \ \int d^4 \pi \, e^{\displaystyle \,i \pi_{\beta}\bar\pi_{\dot\beta} x^{\dot\beta\beta}}\,\pi_\alpha\,
\Psi^{(-1/2)} (\pi,\bar\pi;\xi,\bar\xi)\,,
\end{equation}
which by construction possesses the external spinor index $\alpha$.
This field satisfies the Dirac-Pauli-Fierz equation
\begin{equation}
\label{PF-tw}
\partial^{\dot\alpha\alpha}\,\Phi_\alpha(x;\xi,\bar\xi) \ = \ 0
\end{equation}
and equations \eqref{1eq-tw}, \eqref{2eq-tw}, \eqref{3eq-tw}:
\begin{eqnarray}
\label{1eq-tw-12}
\left(i\xi^{\beta}\partial_{\beta\dot\beta}\bar\xi^{\dot\beta}+M\right)
\Phi_\alpha(x;\xi,\bar\xi) \ = \ 0 \,,
\\ [7pt]
\label{2eq-tw-12}
\left(i\frac{\partial}{\partial\xi_{\beta}}\partial_{\beta\dot\beta}
\frac{\partial}{\partial\bar\xi_{\dot\beta}}-M\right)\Phi_\alpha(x;\xi,\bar\xi) \ = \ 0 \,,
\\ [7pt]
\label{3eq-tw-12}
\left(\xi_{\beta}\frac{\partial}{\partial\xi_{\beta}}-\bar\xi_{\dot\beta}
\frac{\partial}{\partial\bar\xi_{\dot\beta}}\right)\Phi_\alpha(x;\xi,\bar\xi) \ = \ 0 \,.
\end{eqnarray}

The complex conjugate twistorial field \eqref{wf-tw-hel-12c} with charge $c=+1/2$
is used for definition of the spinor field with the dotted Weyl index
\begin{equation}
\label{wf-st-tw-12-d}
\bar\Phi_{\dot\alpha}(x;\xi,\bar\xi) \ = \ \int d^4 \pi \, e^{\displaystyle \,-i \pi_{\beta}\bar\pi_{\dot\beta} x^{\dot\beta\beta}}\,\bar\pi_{\dot\alpha}\,
\bar\Psi^{(+1/2)} (\pi,\bar\pi;\xi,\bar\xi)\,.
\end{equation}
The field \eqref{wf-st-tw-12-d} satisfies the Dirac-Pauli-Fierz equation
\begin{equation}
\label{PF-tw-d}
\partial^{\dot\alpha\alpha}\,\bar\Phi_{\dot\alpha}(x;\xi,\bar\xi) \ = \ 0
\end{equation}
and the equations
\begin{eqnarray}
\label{1eq-tw-12-d}
\left(i\xi^{\beta}\partial_{\beta\dot\beta}\bar\xi^{\dot\beta}-M\right)\bar\Phi_{\dot\alpha}(x;\xi,\bar\xi) \ = \ 0 \,,
\\ [7pt]
\label{2eq-tw-12-d}
\left(i\frac{\partial}{\partial\xi_{\beta}}\partial_{\beta\dot\beta}
\frac{\partial}{\partial\bar\xi_{\dot\beta}}+M\right)\bar\Phi_{\dot\alpha}(x;\xi,\bar\xi) \ = \ 0 \,,
\\ [7pt]
\label{3eq-tw-12-d}
\left(\xi_{\beta}\frac{\partial}{\partial\xi_{\beta}}-\bar\xi_{\dot\beta}
\frac{\partial}{\partial\bar\xi_{\dot\beta}}\right)\bar\Phi_{\dot\alpha}(x;\xi,\bar\xi) \ = \ 0 \,.
\end{eqnarray}

We stress that although the twistorial fields \eqref{wf-tw-hel-12}, \eqref{wf-tw-hel-12c}
have nonvanishing external charges $c=\mp 1/2$, their
integral transformations \eqref{wf-st-tw-12}, \eqref{wf-st-tw-12-d}
have zero $\mathrm{U}(1)$-charge
defined by equations  \eqref{3eq-tw-12}, \eqref{3eq-tw-12-d}. This fact is crucial for forming
infinite spin supermultiplets, as we will see in the next Section.

\setcounter{equation}{0}
\section{Infinite spin supermultiplet}

\quad\,
The results of the previous section allow us to construct an infinite spin supermultiplet.
To do this, we consider the fields $\Phi(x;\xi,\bar\xi)$ and $\Phi_\alpha(x;\xi,\bar\xi)$,
which are defined in \p{wf-st-tw}, \p{wf-st-tw-12}, satisfy equations
\p{KG-tw}-\p{3eq-tw}, \p{PF-tw}-\p{3eq-tw-12} and unify them into one multiplet. These fields contain
the bosonic $\psi^{(k)}(\pi,\bar \pi)$ and fermionic $\psi^{(k-1/2)}(\pi,\bar \pi)$
 component fields ($k \in \mathbb{Z}$) with all
 integer and half-integer spins, respectively.
 It is natural to expect that
individual components of these fields  with fixed spins $n$ and $n+\frac{1}{2}$ ($n=0,1\ldots$) should form the on-shell $\mathcal{N}{=}\,1$
higher spin supermultiplet. Therefore, the
 bosonic (even) $\Phi(x;\xi,\bar\xi)$ and fermionic (odd)
 $\Phi_\alpha(x;\xi,\bar\xi)$ fields themselves should form
the on-shell $\mathcal{N}{=}\,1$ infinite spin supermultiplet containing an infinite number of conventional supermultiplets.
The only thing we should do
is to introduce the corresponding supertrasformations.

Similar to the Wess-Zumino supermultiplet
(see, e.g., \cite{WessB,BK}), we define  supersymmetry transformations for the fields $\Phi$ and $\Phi_\alpha$ in the form
\begin{equation}
\label{tr-susy}
\delta\,\Phi \ = \ \varepsilon^{\alpha}\Phi_\alpha \,, \qquad
\delta\,\Phi_\alpha \ = \ 2i\bar\varepsilon^{\dot\beta}\partial_{\alpha\dot\beta}\Phi \,,
\end{equation}
where $\varepsilon_{\alpha}$, $\bar\varepsilon_{\dot\alpha}$ are the components of the constant odd Weyl spinor.
The commutators of these transformations are
\begin{equation}\label{com-susy}
\begin{array}{lcl}
\left(\delta_1\delta_2-\delta_2\delta_1\right)\Phi &=& -2i a^{\beta\dot\beta}\partial_{\beta\dot\beta}\Phi\,,
\\ [6pt]
\left(\delta_1\delta_2-\delta_2\delta_1\right)\Phi_\alpha &=& -2i a^{\beta\dot\beta}\partial_{\beta\dot\beta}\Phi_\alpha
+2ia_{\alpha\dot\beta}\partial^{\dot\beta\beta}\Phi_\beta\,,
\end{array}
\end{equation}
where
\begin{equation}
\label{a-def}
a_{\alpha\dot\beta} \ := \ \varepsilon_{1\alpha}\bar\varepsilon_{2\dot\beta}- \varepsilon_{2\alpha}\bar\varepsilon_{1\dot\beta}\,.
\end{equation}
As we see, the superalgebra \p{com-susy} is closed on-shell on the translations with the generator
$$
P_{\beta\dot\beta}=-i\partial_{\beta\dot\beta}
$$
due to the equation of motion \p{PF-tw}.
Moreover, the set of equations  \p{KG-tw}-\p{3eq-tw}, \p{PF-tw}-\p{3eq-tw-12} is invariant
with respect to transformations  \p{com-susy}.

Using the integral transformations inverse to \p{wf-st-tw}, \p{wf-st-tw-12},
 we rewrite \p{tr-susy}
as supersymmetry transformations for the twistor fields
$\Psi^{(0)} (\pi,\bar\pi;\xi,\bar\xi)$, $\Psi^{(-1/2)} (\pi,\bar\pi;\xi,\bar\xi)$ in the momentum representation:
\begin{equation}
\label{tr-susy-tw}
\delta\,\Psi^{(0)} \ = \ \varepsilon^{\alpha}\pi_\alpha \Psi^{(-1/2)}\,, \qquad
\delta\,\Psi^{(-1/2)} \ = \ -2\,\bar\varepsilon^{\dot\alpha}\pi_{\dot\alpha}\Psi^{(0)} \,.
\end{equation}
Using further component expansions \p{wf-tw-hel-0}, \p{wf-tw-hel-12}, we find supersymmetry transformations
for the bosonic $A^{(k)}( \pi,\bar \pi) \equiv
\psi^{(k)}( \pi,\bar \pi)$ and fermionic $\psi^{(-\frac12+k)}( \pi,\bar \pi)$ twistorial components
at all $k\,{\in}\,\mathbb{Z}$ in the form
\begin{equation}
\label{tr-susy-tw-comp}
\delta\,A^{(k)} \ = \ \varepsilon^{\alpha}\pi_\alpha \,\psi^{(-\frac12+k)}\,, \qquad
\delta\,\psi^{(-\frac12+k)} \ = \ -2\,\bar\varepsilon^{\dot\alpha}\pi_{\dot\alpha}\,A^{(k)} \,.
\end{equation}
According to \p{const-tw-4-q1-s}, the bosonic field $A^{(k)}$ and fermionic field $\psi^{(-\frac12+k)}$ at fixed
$k\,{\in}\,\mathbb{Z}$ describe massless states with helicities $(-k)$ and $(\frac12-k)$, respectively.
Thus, the infinite-component supermultiplet of the infinite spin stratifies into
an infinite number of levels with pairs of the fields $A^{(k)}$, $\psi^{(-\frac12+k)}$ at fixed $k\,{\in}\,\mathbb{Z}$.
The supersymmetry transforms the bosonic and fermionic
fields into each other inside a given level $k$.
The boosts of the Poincare group transform the levels with
different $k$ and therefore mix the fields with different
values of $k$.

The algebraic structure of the infinite spin supersymmetry was discussed in \cite{BKRX}. Our consideration
is the explicit field description of the supermultiplet introduced in the paper \cite{BKRX}.

\setcounter{equation}{0}
\section{Summary and outlook}

\quad\,
In this paper, we have presented the new twistorial field formulation of the massless infinite spin particle and field.
As opposed to the paper  \cite{BFIR}, we obtained the field description by the canonical quantization of
the world-line twistor model without any gauge fixing.
We gave the helicity decomposition of twistorial infinite spin fields
and constructed the field twistor transform to define
the space-time infinite (continuous) spin fields.
These space-time fields, bosonic $\Phi(x;\xi,\bar\xi)$ and fermionic  $\Phi_\alpha(x;\xi,\bar\xi),$ depend
on the coordinate four-vector $x^m$
and on the commuting Weyl spinors $\xi^\alpha,\,\bar{\xi}^{\dot{\alpha}}$.
We found the equations of motion for $\Phi(x;\xi,\bar\xi)$ and $\Phi_\alpha(x;\xi,\bar\xi)$.
Moreover, we showed that these fields form the $\mathcal{N}{=}\,1$ infinite spin supermultiplet.

It is important to emphasize that the twistor field \p{wf-sol}
(or, equivalently, \p{wf-tw}) was found in the framework of the
scheme of Dirac canonical quantization of one-dimensional gauge
system with constraints \p{L-tw}. First class constraints
\p{M-constr-def}, \p{constr-tw}, \p{const-tw-4} (or
\p{M-constr-def-n}, \p{constr-tw-n}, \p{const-tw-4-n}) yield the
field equations of motion, and it means that the fields of
infinite spin under consideration are determined only on the mass
shell. By construction, these fields do not possess any gauge
symmetry and describe the true physical degrees of freedom. The
space-time fields \p{wf-st-tw}, \p{wf-st-tw-12}, obtained by using the
twistor transform, are also gauge-independent. Of course, a natural
question arises about status of such fields in Lagrangian field
theory. The off-shell Lagrangian field theory can be described in
terms of potentials or field strengths, however on mass shell the
distinctions between potentials and strengths are imperceptible.
Analysis of the Lagrangian field theory will be an aim of our
subsequent works. We expect also that the Lagrangian formulation for
the infinite spin fields should provide additional information on
the role of the index (charge) $c$ presented in the definition of
the twistor fields \p{wf-sol}, \p{wf-tw}.

In subsequent works we will consider the construction of the Lagrangian field
theory of continuous spin, both in the bosonic and fermionic cases and also in the supersymmetric case.
One of the commonly used methods for this purpose is the BRST quantization method, which was used
in the case of continuous spin particles in
\cite{Bengtsson13,AlkGr,Metsaev18,BuchKrTak,ACG18,Metsaev18a}.
In a recent paper \cite{BuchKrTak} the Lagrangian formulation of the infinite integer-spin field
was constructed by using the methods developed in \cite{BuchKrP,BuchKr}.
We plan to construct the Lagrangian formulation for the infinite half-integer field, as well as in the supersymmetric case.
Another interesting problem is to develop the Lagrange description of the infinite spin supermultiplet in a superfield approach.
We note in this regard that in \cite{Zin} the Lagrangian formulation of the infinite spin supermultiplets in the $d{=}\,3$ space-time
was constructed by using the frame-like formalism for corresponding bosonic and fermionic fields.

\section*{Acknowledgments}
The authors are thankful to V.A. Krykhtin for useful discussions.
I.L.B. and A.P.I acknowledge the partial support of the 
Ministry of Science and High Education of Russian Federation,
project No.\,3.1386.2017. I.L.B acknowledges the support of the Russian
Foundation for Basic Research, project No.\,18-02-00153.
The research of S.F. was supported by the Russian Science
Foundation, grant No.\,16-12-10306. A.P.I. acknowledges the support
of the Russian Foundation for Basic Research, grant No.\,19-01-00726.

\section*{Appendix A: Proof of equations  \p{KG-tw}-\p{3eq-tw}, \p{PF-tw}-\p{3eq-tw-12} and \p{PF-tw-d}-\p{3eq-tw-12-d}}
\def\theequation{A.\arabic{equation}}
\setcounter{equation}0

\quad\,
In this Appendix we present the proof of the equations  of motion for the infinite spin space-time fields
\eqref{wf-st-tw}, \eqref{wf-st-tw-12}, \eqref{wf-st-tw-12-d}.

The field \eqref{wf-st-tw} $\Phi(x;\xi,\bar\xi)$ satisfies the massless Klein-Gordon equation  \eqref{KG-tw}
due to resolved form $p_{\alpha\dot\alpha}=\pi_{\alpha}\bar\pi_{\dot\alpha}$ of the momentum in the integrand.

Due to the presence of the $\delta$-function  \eqref{delta-i} in the integrand, the field \eqref{wf-st-tw} satisfies
equations \eqref{1eq-tw}.

The proof of the fulfillment of equation \eqref{2eq-tw} is the following:
\begin{eqnarray}\label{eq-wf-ful2}
&&i\frac{\partial}{\partial\xi_{\alpha}}\partial_{\alpha\dot\alpha}
\frac{\partial}{\partial\bar\xi_{\dot\alpha}} \ \Phi  =
\\
&&\int d^4 \pi \, e^{\displaystyle \,i \pi_{\alpha}\bar\pi_{\dot\alpha} x^{\dot\alpha\alpha}}\,
\delta\left((\pi\xi)(\bar\xi\bar\pi)-M\right)
\left[-\pi_\alpha\frac{\partial}{\partial\xi_{\alpha}}\bar\pi_{\dot\alpha}
\frac{\partial}{\partial\bar\xi_{\dot\alpha}}e^{\displaystyle -iq_0/p_0}\right]
\hat\Psi^{(0)} (\pi,\bar \pi;\xi,\bar\xi)  = M\,\Phi \,. \nonumber
\end{eqnarray}

The proof of fulfillment of equation \eqref{3eq-tw} is the following:
\begin{eqnarray}\label{eq-wf-ful3}
&&\left(\xi_{\alpha}\frac{\partial}{\partial\xi_{\alpha}}-\bar\xi_{\dot\alpha}
\frac{\partial}{\partial\bar\xi_{\dot\alpha}}\right) \Phi  = \\
&&\int d^4 \pi \, e^{\displaystyle \,i \pi_{\alpha}\bar\pi_{\dot\alpha} x^{\dot\alpha\alpha}}\,
\delta\left((\pi\xi)(\bar\xi\bar\pi)-M\right)
\left[\left(\xi_{\alpha}\frac{\partial}{\partial\xi_{\alpha}}-\bar\xi_{\dot\alpha}
\frac{\partial}{\partial\bar\xi_{\dot\alpha}}\right)e^{\displaystyle -iq_0/p_0}\right]
\hat\Psi^{(0)} (\pi,\bar \pi;\xi,\bar\xi)  + \nonumber \\
&&\qquad\int d^4 \pi \, e^{\displaystyle \,i \pi_{\alpha}\bar\pi_{\dot\alpha} x^{\dot\alpha\alpha}}\,
\delta\left((\pi\xi)(\bar\xi\bar\pi)-M\right)
e^{\displaystyle -iq_0/p_0}\left(\xi_{\alpha}\frac{\partial}{\partial\xi_{\alpha}}-\bar\xi_{\dot\alpha}
\frac{\partial}{\partial\bar\xi_{\dot\alpha}}\right)
\hat\Psi^{(0)} (\pi,\bar \pi;\xi,\bar\xi)  = \nonumber \\
&&\int d^4 \pi \, e^{\displaystyle \,i \pi_{\alpha}\bar\pi_{\dot\alpha} x^{\dot\alpha\alpha}}\,
\delta\left((\pi\xi)(\bar\xi\bar\pi)-M\right)
\left[-\left(\pi_{\alpha}\frac{\partial}{\partial\pi_{\alpha}}-\bar\pi_{\dot\alpha}
\frac{\partial}{\partial\bar\pi_{\dot\alpha}}\right)e^{\displaystyle -iq_0/p_0}\right]
\hat\Psi^{(0)} (\pi,\bar \pi;\xi,\bar\xi)  + \nonumber \\
&&\qquad\int d^4 \pi \, e^{\displaystyle \,i \pi_{\alpha}\bar\pi_{\dot\alpha} x^{\dot\alpha\alpha}}\,
\delta\left((\pi\xi)(\bar\xi\bar\pi)-M\right)
e^{\displaystyle -iq_0/p_0}\left(\xi_{\alpha}\frac{\partial}{\partial\xi_{\alpha}}-\bar\xi_{\dot\alpha}
\frac{\partial}{\partial\bar\xi_{\dot\alpha}}\right)
\hat\Psi^{(0)} (\pi,\bar \pi;\xi,\bar\xi)  = \nonumber \\
&&\int d^4 \pi \, e^{\displaystyle \,i \pi_{\alpha}\bar\pi_{\dot\alpha} x^{\dot\alpha\alpha}}\,
\delta\left((\pi\xi)(\bar\xi\bar\pi)-M\right)
e^{\displaystyle -iq_0/p_0}\,\cdot
\nonumber \\
&&\qquad\qquad\qquad\qquad\qquad \cdot\left(\pi_{\alpha}\frac{\partial}{\partial\pi_{\alpha}}-\bar\pi_{\dot\alpha}
\frac{\partial}{\partial\bar\pi_{\dot\alpha}}+\xi_{\alpha}\frac{\partial}{\partial\xi_{\alpha}}-\bar\xi_{\dot\alpha}
\frac{\partial}{\partial\bar\xi_{\dot\alpha}}\right)
\hat\Psi^{(0)} (\pi,\bar \pi;\xi,\bar\xi) \ = \ 0 \,.\nonumber
\end{eqnarray}

The spinor field $\Phi_\alpha(x;\xi,\bar\xi)$ defined by integral transform \eqref{wf-st-tw-12}
satisfies the massless Dirac equation  \p{PF-tw}
due to $p_{\alpha\dot\alpha}=\pi_{\alpha}\bar\pi_{\dot\alpha}$ in integrand and the identity
 $\pi^\alpha \pi_\alpha \equiv 0$
for the even Weyl spinor. The proof of fulfillment of equations \eqref{1eq-tw-12}, \eqref{2eq-tw-12} for the spinor field
$\Phi_\alpha(x;\xi,\bar\xi)$ is completely analogous to the scalar field $\Phi (x;\xi,\bar\xi)$.
The proof of fulfillment of equation \p{3eq-tw-12} is some generalization of \p{eq-wf-ful3}:
\begin{eqnarray}\label{eq-wf-ful3-s}
&&\left(\xi_{\alpha}\frac{\partial}{\partial\xi_{\alpha}}-\bar\xi_{\dot\alpha}
\frac{\partial}{\partial\bar\xi_{\dot\alpha}}\right) \Phi_\beta  = \\
&&\int d^4 \pi \, e^{\displaystyle \,i \pi_{\alpha}\bar\pi_{\dot\alpha} x^{\dot\alpha\alpha}}\,
\pi_\beta\, \delta\left((\pi\xi)(\bar\xi\bar\pi)-M\right)
\left[\left(\xi_{\alpha}\frac{\partial}{\partial\xi_{\alpha}}-\bar\xi_{\dot\alpha}
\frac{\partial}{\partial\bar\xi_{\dot\alpha}}\right)e^{\displaystyle -iq_0/p_0}\right]
\hat\Psi^{(-1/2)} (\pi,\bar \pi;\xi,\bar\xi)  + \nonumber \\
&&\qquad\int d^4 \pi \, e^{\displaystyle \,i \pi_{\alpha}\bar\pi_{\dot\alpha} x^{\dot\alpha\alpha}}\,
\pi_\beta\, \delta\left((\pi\xi)(\bar\xi\bar\pi)-M\right)
e^{\displaystyle -iq_0/p_0}\left(\xi_{\alpha}\frac{\partial}{\partial\xi_{\alpha}}-\bar\xi_{\dot\alpha}
\frac{\partial}{\partial\bar\xi_{\dot\alpha}}\right)
\hat\Psi^{(-1/2)} (\pi,\bar \pi;\xi,\bar\xi)  = \nonumber \\
&&\int d^4 \pi \, e^{\displaystyle \,i \pi_{\alpha}\bar\pi_{\dot\alpha} x^{\dot\alpha\alpha}}\,
\pi_\beta\, \delta\left((\pi\xi)(\bar\xi\bar\pi)-M\right)
\left[-\left(\pi_{\alpha}\frac{\partial}{\partial\pi_{\alpha}}-\bar\pi_{\dot\alpha}
\frac{\partial}{\partial\bar\pi_{\dot\alpha}}\right)e^{\displaystyle -iq_0/p_0}\right]
\hat\Psi^{(-1/2)} (\pi,\bar \pi;\xi,\bar\xi)  + \nonumber \\
&&\qquad\int d^4 \pi \, e^{\displaystyle \,i \pi_{\alpha}\bar\pi_{\dot\alpha} x^{\dot\alpha\alpha}}\,
\pi_\beta\, \delta\left((\pi\xi)(\bar\xi\bar\pi)-M\right)
e^{\displaystyle -iq_0/p_0}\left(\xi_{\alpha}\frac{\partial}{\partial\xi_{\alpha}}-\bar\xi_{\dot\alpha}
\frac{\partial}{\partial\bar\xi_{\dot\alpha}}\right)
\hat\Psi^{(-1/2)} (\pi,\bar \pi;\xi,\bar\xi)  = \nonumber \\
&&\int d^4 \pi \, e^{\displaystyle \,i \pi_{\alpha}\bar\pi_{\dot\alpha} x^{\dot\alpha\alpha}}\,
\pi_\beta\, \delta\left((\pi\xi)(\bar\xi\bar\pi)-M\right)
e^{\displaystyle -iq_0/p_0}\,\cdot
\nonumber \\
&&\qquad\qquad\qquad\qquad\qquad \cdot\left(\pi_{\alpha}\frac{\partial}{\partial\pi_{\alpha}}-\bar\pi_{\dot\alpha}
\frac{\partial}{\partial\bar\pi_{\dot\alpha}}+\xi_{\alpha}\frac{\partial}{\partial\xi_{\alpha}}-\bar\xi_{\dot\alpha}
\frac{\partial}{\partial\bar\xi_{\dot\alpha}}+1\right)
\hat\Psi^{(-1/2)} (\pi,\bar \pi;\xi,\bar\xi) \ = \ 0 \,.\nonumber
\end{eqnarray}

Since $\bar\Phi_{\dot\alpha}(x;\xi,\bar\xi)$ is complex conjugation of the field $\Phi_{\alpha}(x;\xi,\bar\xi)$,
therefore the fulfilment of equations \eqref{PF-tw-d}, \eqref{1eq-tw-12-d}, \eqref{2eq-tw-12-d}, \eqref{3eq-tw-12-d} is the
consequence of equations \eqref{PF-tw}, \eqref{1eq-tw-12}, \eqref{2eq-tw-12}, \eqref{3eq-tw-12}.

\section*{Appendix B: Momentum wave function}
\def\theequation{B.\arabic{equation}}
\setcounter{equation}0

\quad\,\,In this Appendix we present  the exact form of the momentum wave function.

For definiteness we consider the case  $c=0$.

In integrand \eqref{wf-st-tw} the space-time momentum  is represented as a product of
spinor components \p{p-resol}: $p_{\alpha\dot\alpha}=\pi_{\alpha}\bar\pi_{\dot\alpha}$.
Inverse expressions for the spinor $\pi_{\alpha}$ are
\begin{equation}
\label{pi-pi}
\pi_{\alpha}=\bm{\pi}_{\alpha}(p)\,e^{i\varphi}\,,\qquad
\bar\pi_{\dot\alpha}=\bar{\bm{\pi}}_{\dot\alpha}(p)\,e^{-i\varphi}\,,
\end{equation}
where
\begin{equation}
\label{pi-p}
\bm{\pi}_{\alpha}(p):\,\,\bm{\pi}_{1}=|\pi_1|e^{i\phi}\,,\,\,\bm{\pi}_{2}=|\pi_2|e^{-i\phi}\,,\qquad
\bar{\bm{\pi}}_{\dot\alpha}(p):\,\,\bar{\bm{\pi}}_{\dot 1}=|\pi_1|e^{-i\phi}\,,\,\,\bar{\bm{\pi}}_{\dot 2}=|\pi_2|e^{i\phi}
\end{equation}
are the functions of the light-like momentum components:
\begin{equation}
\label{pi-p1}
|\pi_1|= \left(\frac{\left|p^0+p^3 \right|}{\sqrt{2}}\right)^{1/2}\,,\qquad
|\pi_2|= \left(\frac{\left|p^0-p^3 \right|}{\sqrt{2}}\right)^{1/2}\,,\qquad
e^{i\phi}=\left(\frac{p^1-ip^2}{p^1+ip^2} \right)^{1/4}\,.
\end{equation}
The phase $\varphi$ presented in \p{pi-pi} does not give a contribution to the definition
of the light-like momentum $p_{\alpha\dot\alpha}=\bm{\pi}_{\alpha}\bar{\bm{\pi}}_{\dot\alpha}$.
As result, we have
\begin{equation}
\label{measure-p1}
d^4 \pi = \frac18\, d\pi^{\alpha}\wedge d\pi_{\alpha}\wedge d\bar\pi^{\dot\alpha}\wedge d\bar\pi_{\dot\alpha}=d^4p\,\delta(p^2)\,d\varphi\,.
\end{equation}
In terms of the variables \p{pi-pi} the fraction \p{w0-p0} takes the value
\begin{equation}
\label{w0-p0-a}
\frac{q_0}{p_0} \ = \
a(p;\xi,\bar\xi)\cos\varphi\ +\ b(p;\xi,\bar\xi)\sin\varphi \,,
\end{equation}
where
\begin{equation}
\label{ab-def}
a(p;\xi,\bar\xi) \ = \ \frac{\sqrt{M}\sum\limits_{\alpha=\dot\alpha}({\bm{\pi}}_{\alpha}\bar\xi_{\dot\alpha} + \xi_{\alpha}\bar{\bm{\pi}}_{\dot\alpha})}
{\sum\limits_{\beta=\dot\beta}{\bm{\pi}}_{\beta}\bar{\bm{\pi}}_{\dot\beta}} \,,\qquad
b(p;\xi,\bar\xi) \ = \ \frac{i\sqrt{M}\sum\limits_{\alpha=\dot\alpha}({\bm{\pi}}_{\alpha}\bar\xi_{\dot\alpha} - \xi_{\alpha}\bar{\bm{\pi}}_{\dot\alpha})}
{\sum\limits_{\beta=\dot\beta}{\bm{\pi}}_{\beta}\bar{\bm{\pi}}_{\dot\beta}} \,.
\end{equation}
In addition, equations  \eqref{const-tw-4-q1-n} for
the component fields $\psi^{(k)}(\pi,\bar \pi)$ of the wave function \eqref{wf-tw-hel-0} take the form
(at $c=0$):
\begin{equation}
\label{const-tw-4-q1-n-i-ap}
\left(\pi_{\alpha}\frac{\partial}{\partial \pi_{\alpha}} \ -\
\bar \pi_{\dot\alpha}\frac{\partial}{\partial \bar \pi_{\dot \alpha}}\right)
\psi^{(k)}( \pi,\bar \pi)\ =\  -i\,\frac{\partial}{\partial \varphi} \,
\psi^{(k)}( \pi,\bar \pi)\ =\  2k\, \psi^{(k)}( \pi,\bar \pi)\,.
\end{equation}
General dependence of the component fields $\psi^{(k)}(\pi,\bar \pi)$ on the phase $\varphi$
is defined by the Fourier series. Then, general solutions of equations \eqref{const-tw-4-q1-n-i-ap} are
\begin{equation}
\label{sol-psi-ap}
\psi^{(k)}( \pi,\bar \pi)\ =\ e^{2ik\varphi}\,{\bm{\psi}}^{(k)}( {\bm{\pi}},\bar {\bm{\pi}})\ = \ e^{2ik\varphi}\,{\bm{\psi}}^{(k)}(p) \,,
\end{equation}
where the functions ${\bm{\psi}}^{(k)}( p)$ do not depend on $\varphi$.

Inserting \p{pi-pi}, \p{measure-p1}, \p{w0-p0-a}, \p{ab-def}, \p{sol-psi-ap}
in \p{wf-tw-hel-0} and  \eqref{wf-st-tw} we obtain that the space-time field of the infinite spin particle takes the form
\begin{equation}
\label{wf-st-tw-ap}
\Phi(x;\xi,\bar\xi) \ = \ \int d^4 p\,e^{\displaystyle \,i p\, x}\,
{\bm{\Psi}}(p,\xi,\bar\xi)\,,
\end{equation}
where
\begin{eqnarray}
\label{wf-tw-hel1-ap}
{\bm{\Psi}}(p,\xi,\bar\xi) &=& \delta(p^2) \,
\delta(\xi^{\alpha}p_{\alpha\dot\alpha}\bar\xi^{\dot\alpha}-M)
\left[\,{\bm{\psi}}^{(0)}(p)\int\limits_{0}^{2\pi}d\varphi\, e^{\displaystyle -i\left(a\cos\varphi + b\sin\varphi \right)}\right.
\\ [6pt]
\nonumber
&&
+
\sum\limits_{k=1}^{\infty} {\bm{\psi}}^{(k)}(p)\,(\bar\xi\bar{\bm{\pi}})^{k}
\int\limits_{0}^{2\pi}d\varphi\, e^{\displaystyle i\left(k\varphi -a\cos\varphi - b\sin\varphi \right)}
\\ [6pt]
\nonumber
&&
\left.  +
\sum\limits_{k=1}^{\infty} {\bm{\psi}}^{(-k)}(p)\,({\bm{\pi}}\xi)^{k}
\int\limits_{0}^{2\pi}d\varphi\, e^{\displaystyle -i\left(k\varphi +a\cos\varphi + b\sin\varphi \right)}\right].
\end{eqnarray}

We will use the integral (see formulae\,337\,(9a,b) in \cite{GrHof} and formulae\,3.937\,(1,2) in \cite{GR})
\begin{equation}
\label{int-GR}
\int\limits_0^{2\pi} e^{\displaystyle i\left(\pm n\varphi-a\cos\varphi-b\sin\varphi \right)}d\varphi \ = \ 2\pi
\left(\frac{\displaystyle a\pm ib}{\displaystyle a\mp ib} \right)^{n/2} I_n(i\sqrt{a^2+b^2})\,,
\end{equation}
where $n=0,1,2,\ldots$ and $I_n$ are the modified Bessel functions.\footnote{ Modified Bessel functions
(or Bessel functions of imaginary argument) $I_n(z)=I_{-n}(z)$ are expressed in terms of the Bessel functions of the first kind $J_n(z)$ by the expression
\begin{equation}
\label{B-mB}
I_n(z)=i^{-n}J_n(iz)
\end{equation}
in the cases when $n$ is integer. }

Expressions \p{ab-def} lead to the following equalities:
\begin{equation}
\label{a-b-eq}
a+ib \ = \ \frac{2\sqrt{M}\sum\limits_{\alpha=\dot\alpha}\xi_{\alpha}\bar{\bm{\pi}}_{\dot\alpha}}
{\sum\limits_{\beta=\dot\beta}{\bm{\pi}}_{\beta}\bar{\bm{\pi}}_{\dot\beta}} \,,\qquad
a-ib \ = \ \frac{2\sqrt{M}\sum\limits_{\alpha=\dot\alpha}{\bm{\pi}}_{\alpha}\bar\xi_{\dot\alpha} }
{\sum\limits_{\beta=\dot\beta}{\bm{\pi}}_{\beta}\bar{\bm{\pi}}_{\dot\beta}} \,.
\end{equation}
Therefore, there are equalities
\begin{equation}
\label{a-b-eq1}
\frac{a+ib}{a-ib} \ = \ \frac{\sum\limits_{\alpha=\dot\alpha}\xi_{\alpha}\bar{\bm{\pi}}_{\dot\alpha}}
{\sum\limits_{\beta=\dot\beta}{\bm{\pi}}_{\beta}\bar\xi_{\dot\beta}} \,,
\end{equation}
\begin{equation}
\label{a-b-eq2}
y^2 \ := \ a^2+b^2 \ = \ \frac{4M\,\sum\limits_{\alpha=\dot\alpha}\xi_{\alpha}\bar{\bm{\pi}}_{\dot\alpha}\,
\sum\limits_{\beta=\dot\beta}{\bm{\pi}}_{\beta}\bar\xi_{\dot\beta}}
{\big(\sum\limits_{\gamma=\dot\gamma}{\bm{\pi}}_{\gamma}\bar{\bm{\pi}}_{\dot\gamma}\big)^2} \,.
\end{equation}

Using \p{int-GR}-\p{a-b-eq2}, we obtain that the momentum wave function \p{wf-tw-hel1-ap} takes the form
\begin{eqnarray}
\label{wf-tw-hel1-ap1}
{\bm{\Psi}}(p,\xi,\bar\xi) &=& 2\pi\,\delta(p^2) \,
\delta(\xi^{\alpha}p_{\alpha\dot\alpha}\bar\xi^{\dot\alpha}-M)
\Bigg[\,{\bm{\psi}}^{(0)}(p)I_0(i y)
\\ [6pt]
\nonumber
&&
+
\sum\limits_{k=1}^{\infty} {\bm{\psi}}^{(k)}(p)\,\left(\frac{(\bar\xi\bar{\bm{\pi}})^{2}\sum\limits_{\alpha=\dot\alpha}\xi_{\alpha}\bar{\bm{\pi}}_{\dot\alpha}}
{\sum\limits_{\beta=\dot\beta}{\bm{\pi}}_{\beta}\bar\xi_{\dot\beta}} \right)^{k/2}I_k(i y)
\\ [6pt]
\nonumber
&&
+
\sum\limits_{k=1}^{\infty} {\bm{\psi}}^{(-k)}(p)
\,\left(\frac{({\bm{\pi}}\xi)^{2}\sum\limits_{\beta=\dot\beta}{\bm{\pi}}_{\beta}\bar\xi_{\dot\beta}}
{\sum\limits_{\alpha=\dot\alpha}\xi_{\alpha}\bar{\bm{\pi}}_{\dot\alpha}} \right)^{k/2} I_k(i y)\Bigg]\,,
\end{eqnarray}
where ${\bm{\psi}}^{(k)}(p)$ at fixed $k\in \mathbb{Z}$ describes the massless state of helicity $-k$.

\end{document}